\documentclass[preprint1]{aastex}
\usepackage{graphicx}
\usepackage{epsfig}
\usepackage{amssymb,amsmath}

\slugcomment{To appear in ApJ}
\shorttitle{The Nature of  Transition Disks}
\shortauthors{Cieza et al.}

\begin{document}

\title{The Nature of  Transition Circumstellar Disks III. \\  Perseus, Taurus, and Auriga.{\LARGE{$^{\star}$}}}

\author{Lucas A. Cieza\altaffilmark{1,}\altaffilmark{2},
Matthias R. Schreiber\altaffilmark{3},
Gisela A. Romero\altaffilmark{3,4},
Jonathan P.  Williams\altaffilmark{1},
Alberto  Rebassa-Mansergas\altaffilmark{3}, and
Bruno Mer\'{i}n\altaffilmark{5}
}

\altaffiltext{}{$\star$ Based in part on observations made with the CFHT, under programs 09BH48, 09BH10, 10AH06, and 10BH97.}
\altaffiltext{1}{Institute for Astronomy, University of Hawaii at Manoa,  Honolulu, HI 96822.} 
\altaffiltext{2}{\emph{Sagan} Fellow, lcieza@ifa.hawaii.edu}
\altaffiltext{3}{Departamento de Fisica y Astronomia, Universidad de Valpara\'{\i}so, Valpara\'{\i}so, Chile}
\altaffiltext{4}{Facultad  de Ciencias Astron\'omicas y Geof\'{\i}sicas, UNLP, La Plata, Argentina}
\altaffiltext{5}{Herschel Science Centre, ESAC-ESA, P.O. Box, 78, 28691 Villanueva de la Ca\~{n}ada, Madrid, Spain}
\begin{abstract}

As part of an ongoing program aiming to characterize a large number
of  \emph{Spitzer}-selected transition disks (disks with reduced levels of near-IR and/or mid-IR excess emission),  
we have obtained (sub)millimeter  wavelength photometry, high-resolution optical spectroscopy,  and adaptive 
optics near-infrared imaging for a sample of  31  transition objects located in the Perseus, Taurus, and Auriga 
molecular clouds.
We use these ground-based data to estimate  disk masses,  
multiplicity, and accretion rates in order to  investigate the mechanisms 
potentially responsible for  their inner holes.
Following our previous studies in other regions,  we combine disk masses, accretion rates
and multiplicity data with other information, such as SED morphology and fractional disk luminosity
to classify the disks as \emph{strong candidates} for the following categories:
grain-growth dominated disks (7 objects), giant planet-forming disks (6 objects), 
photoevaporating disks (7 objects), debris disks (11 objects), and cicumbinary disks 
(1 object,  which was also classified as a photoeavaporating disk). 
Combining our sample of 31 transition disks with those from our previous studies
results in a sample of 74 transition objects that have been selected, characterized, and
classified in an homogenous way.
We discuss  this combined high-quality sample in the context of the current paradigm of 
the evolution and dissipation of protoplanetary disks and use its  properties to constrain 
different aspects of  the key processes driving their evolution.
We  find that the age distribution of disks that are likely to harbor recently formed 
giant planets favors core accretion as the main planet formation mechanism and
 a $\sim$2-3 Myr  formation  timescale.
\end{abstract}

\section{Introduction}\label{intro}

Protoplanetary disks evolve through a variety of physical processes. 
Early in its evolution, a massive gas-rich disk loses mass through 
accretion onto the star, outflows,  and  photevaporation  by high-energy  photons (Gorti et al. 2009). 
At the same time,  grains grow into larger bodies that settle onto  the mid-plane of the disk 
where they can grow into rocks, planetesimals, and beyond.
Dust settling steepens the slope of  the mid- and far-infrared (IR)  Spectral Energy Distribution (SED) because a smaller 
fraction of the stellar radiation is intercepted by circumstellar dust (Dullemond $\&$ Dominik, 2004). 
Also, since grain growth is expected to proceed faster in the inner disk, it could result in an inner 
opacity hole (Dullemond $\&$ Dominik, 2005). 
As the disk mass and accretion rate decrease, chromospheric Extreme-Ultraviolet (EUV) photons 
start to penetrate the inner disk and EUV-induced photoevaporation becomes 
important.  
When the accretion rate drops below the photoevaporation rate, the outer 
disk is no longer able to resupply the inner disk with material.  
At this point,  the inner disk drains on a viscous timescale 
($\sim$10$^5$ yr) 
and an inner hole is formed (Alexander et al. 2006).  Once this inner hole has formed,  
the EUV photons can reach the inner edge of the disk unimpeded,
preventing any material from the outer disk from flowing into the inner hole. 
This halts accretion and results on the rapid transition between the
classical T Tauri star  (CTTS)  
and the  weak-line T Tauri star (WTTS) stage.
The disk then quickly photoevaporates from the inside out. 
Once the  remaining gas photoevaporates,  the small grains are removed by
radiation pressure and Poynting-Robertson  drag.  
What is left constitutes the initial conditions of a debris disk: a gas poor disk with large grains, 
planetesimals and/or planets.

This  evolutionary path is certainly not unique as some systems  such as  LkCa 15,  DM Tau, and GM Aur
seem to have developed sharp cavities due to the dynamical interactions with close (sub)stellar companions or 
recently formed giant planets while their outer disks are still  quite massive 
(Najita et al. 2007).
Even though the broadest aspects of disk evolution summarized above 
are relatively well established (see Williams $\&$ Cieza, 2011 for a recent review on protoplanetary
disks and their evolution),  we are still far from developing a comprehensive  disk evolution theory, 
for which additional observational constraints are much needed.  
The so-called ``transition" objects  (broadly defined as disks with inner opacity holes) are particularly useful
\textit{disk evolution laboratories} as they are the systems where the  key physical processes 
mentioned above have the clearest observational  signatures. 
This is so simply because grain growth, photevaporation, and dynamical clearing all
result in reduced levels of near- and/or mid-IR excess emission, 
which is the defining feature of  transition objects. 

This paper is the third  part of a series from  an ongoing project aiming to characterize 
$\sim$100  \emph{Spitzer}-selected transition disks located in nearby star-forming regions. 
This coordinated project has two main goals: 1) provide observational constraints on
the evolution of primordial disks, their dissipation, and the primordial to debris disk transition,
and 2) identify systems with strong evidence for ongoing giant planet formation so they can eventually 
be used as \emph{planet formation laboratories}. 
In the first paper of the series (Cieza et al. 2010; Paper I hereafter),  we studied a sample of  26 Ophiuchus transition
disks, while in the second paper (Romero et al. 2012, ApJ, in press; Paper II hereafter) we analyzed a sample of  
17 objects from the Lupus, Corona Australis, and Scorpius regions. 
In Papers I and II, we showed that transition disks are a very heterogenous group of objects 
with disk masses ranging from  $<$~0.6 M$_{JUP}$ to 40 M$_{JUP}$, accretion
 rates ranging from $<$~10$^{-11}$ to 10$^{-7}$ M$_{\odot}yr^{-1}$, 
and fractional disk luminosities, L$_{D}$/L$_{*}$,  ranging from a few percent to
$\lesssim$~10$^{-4}$. We also showed that their diverse properties can be understood in
terms of different disk evolution processes and distinct evolutionary stages.

Here we present millimeter wavelength photometry (from the Submillimeter Array and JCMT),  
high-resolution optical spectroscopy (from the Canada-Franca-Hawaii Telescope),  and Adaptive Optics  near-infrared imaging  
(from the Gemini telscope)  for other  31 \emph{Spitzer}-selected transition circumstellar disks  located in the
Perseus, Taurus, and Auriga molecular clouds. 
As in our previous studies, we use these new ground-based data to estimate  disk masses,  accretion rates,
and multiplicity for our sample
in order to  investigate the mechanisms  potentially responsible for their inner opacity holes.  
These new 31 objects take our total sample of  well characterized transition disks to 74.
The structure of this paper is as follows.  Our  sample selection criteria are presented in Section~2, while our 
observations and data reduction techniques are described in Section~3. We present our results on disk masses, 
accretion rates, and multiplicity in Section~4. In Section~5, we discuss the likely origins of the inner holes
of  the 31 new individual targets and discuss the properties of our combined sample of 74 transition 
objects in a broader context of disk evolution. Finally, a summary of our 
results and conclusions is presented in Section~6.

\section{Sample Selection}\label{selection}

\begin{figure}
\includegraphics[width=4.5in]{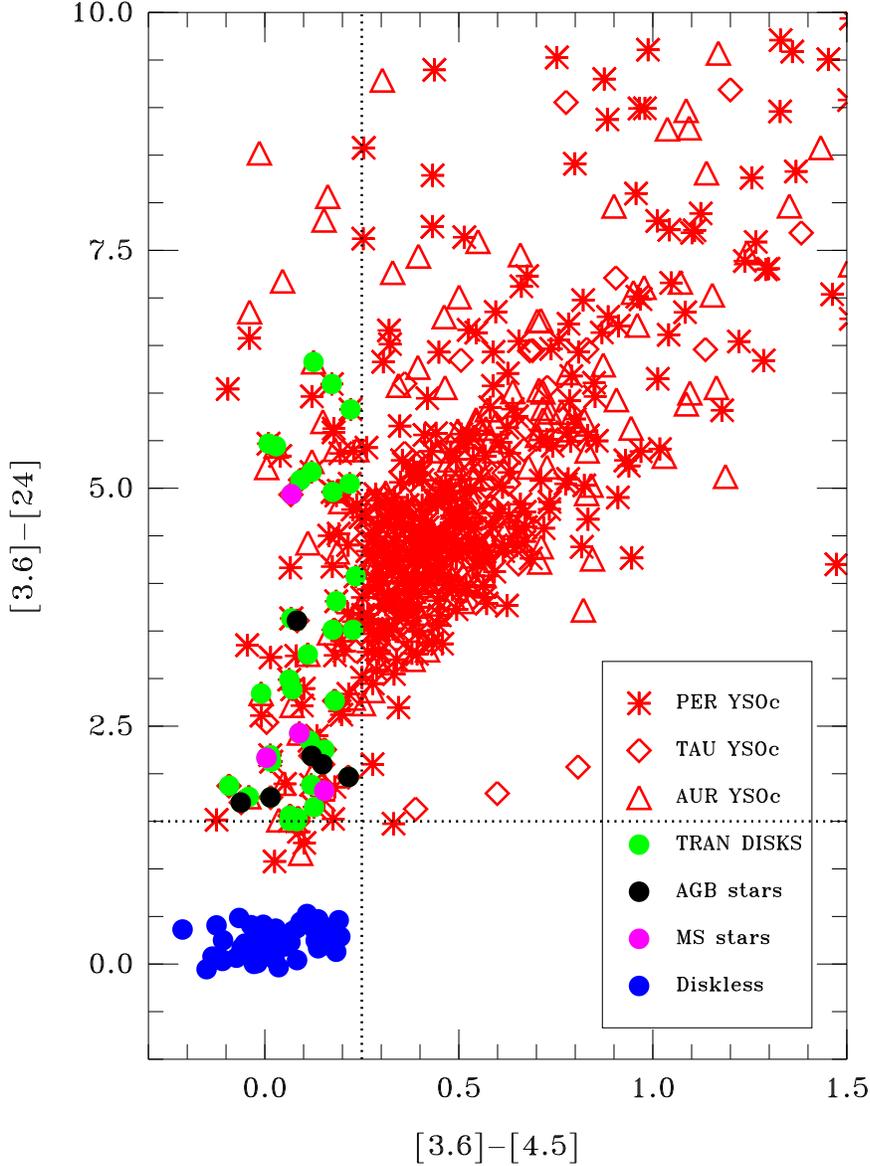}
\caption{
Color-color diagram  illustrating our key target selection criteria.
Objects with [3.6]-[4.5] $<$ 0.25 and [3.6] -[24] $<$ 0.5 are consistent
with bare stellar photospheres. Blue dots are disk-less WTTSs from
Cieza et al. (2007) used to define this region of the diagram.
Red stars and triangles are all the Young Stellar Objects Candidates (YSOc) in the Perseus and 
Auriga catalogs from the \emph{c2d} and \emph{Gould Belt} Legacy Projects.
Since the catalogs from the \emph{Taurus} project  does not provide a YSOc classification, we simply plot all
the targets with IR excesses and 3.6 and 24 $\mu$m fluxes at least as large as those of the faintest transition disks 
in our sample (red diamonds). Most PMS stars are either disk-less or have excesses at both 4.5 and 24 $\mu$m.
Our 31 transition disks, shown as green dots,   have significant ($>$ 5-$\sigma$) excess
at 24 $\mu$m and little or no excess at 4.5 $\mu$m,  as expected for disks with inner holes. 
}
\label{f:sample_sel}
\end{figure}

We drew our sample from the \emph{Spitzer} catalogs of 3 northern clouds mapped
by 3 different Legacy Programs:  \emph{Cores to Disks}  (Perseus molecular cloud\footnote{available through IRSA at  \tt
http://irsa.ipac.caltech.edu/data/SPITZER/C2D/}), 
 \emph{Taurus} (Taurus molecular cloud\footnote{available through IRSA at  \tt http://irsa.ipac.caltech.edu/data/SPITZER/Taurus/}),  
 and \emph{Gould's Belt} (Auriga molecular cloud).
For a description of the \emph{Cores to Disks}  data products, 
see Evans et al.\footnote{available at \tt http://irsa.ipac.caltech.edu/data/SPITZER/C2D/doc/c2d$\_$del$\_$document.pdf}.
The \emph{Taurus} catalogs are discussed by Padgett et al.\footnote{available at \tt  http://irsa.ipac.caltech.edu/data/SPITZER/Taurus/docs/delivery\_doc2.pdf}.
A the time of this writing, the  \emph{Gould's Belt} data products  have not yet been delivered to NASA's Infra-Red Science Archive (IRSA), but they should also 
eventually become publicly available through IRSA.  
As in Papers I and II,  we selected all the targets  meeting the following criteria:  
 
\noindent{\emph{a.}}   Have \emph{Spitzer} colors [3.6]-[4.5]  $<$  0.25.  These are objects with small or 
no near-IR excess (see Figure~\ref{f:sample_sel}). 
The lack of a  [3.6] - [4.5]  color excess in our targets is 
inconsistent with an optically thick  disk extending inward to the dust sublimation radius, and therefore 
implies the presence of an inner opacity hole. 
The presence of this inner opacity hole is the defining feature we  intend to capture in our sample.  

\noindent{\emph{b.}}  Have \emph{Spitzer} colors  [3.6]-[24] $>$ 1.5.  We apply this criterion to ensure that  all our targets 
have very significant 24 $\mu$m excesses ($>$5-10 $\sigma$), unambiguously indicating the presence of
circumstellar material.  The exact color cut is somewhat arbitrary. We have empirically found that  most
YSOc  with low 24 $\mu$m excesses ( [3.6]-[24] $\lesssim$ 1.5) are in fact background AGB stars (see Paper II). 

\noindent{\emph{c.}}   Are detected with a signal to noise ratio $>$ 7 in all  2MASS and  IRAC wavelengths as 
well as at 24 $\mu$m, to ensure that we only include targets with very reliable photometry. 

\noindent{\emph{d}}.   Have K$_S$ $<$ 11 mag ,  driven by the sensitivity of our near-IR  adaptive optics observations 
and to ensure a negligible extragalactic contamination (Padgett et al. 2008). 

\noindent{\emph{e.}}   Are brighter than R  = 18 mag according to the USNO-B1 catalog  (Monet et al. 2003),
driven by the sensitivity of our optical spectroscopy observations.  

The first  two selection criteria ( [3.6]-[4.5]  $<$  0.25 and [3.6]-[24] $>$ 1.5) effectively become  our
working definition for a transition disk. These criteria are fairly inclusive and encompass most of 
the transition disk definitions found in the literature  as they select disks  with a significant  flux decrement  
relative to  ``normal disks"  in the near-IR or at all  wavelengths. In particular, our definition includes objects
with falling mid-IR SEDs such as anemic  (Lada et al. 2006) or homologously depleted transition disks (Currie et la. 2009)  
as well as objects with rising mid-IR SEDs such as classical transition disks 
(Muzerolle et al. 2010) or cold disks (Brown et al. 2007; Merin et al. 2010). 
The one type of transition disk that is likely to be under-represented in our sample is the so-called ``pre-transitional" disk 
category, which describes systems with optically thin gaps separating optically thick inner and outer disk
components  (Espaillat et al. 2007). These rare objects tend to have  large near-IR excesses that 
could be excluded by  our  [3.6]-[4.5]  $<$  0.25 criterion. 
For a concise description of the complex transition disk nomenclature, see Evans et al. (2009).
We find 41 targets that meet all of our selection criteria: 18 objects in Perseus ($\sim$44$\%$), 
6 in Auriga ($\sim$15$\%$), and 17 in Taurus ($\sim$41$\%$).  
We have observed all 41 objects in our target list; however, as discussed in Section~\ref{pms_id}, 
this list includes 
6  Asymptotic Giant Branch (AGB) stars and 4 likely debris disks around main-sequence (MS) stars.
The remaining 31 targets are bona fide young stellar objects  (YSOs) with circumstellar disks and 
constitute our science sample. 
The 2MASS IDs and alternative names, 2MASS and \emph{Spitzer} fluxes, and  the USNO-B1 R-band
magnitudes for all our 41 targets are listed in Table 1.

The Spitzer data of our Perseus transition disks have been presented by
Jorgensen et al. (2006) and Rebull et al. (2007). Similarly, the 
parent sample of our Taurus objects has been discussed by Rebull et al. (2010)
and Luhman et al. (2010). We note that while Rebull et al.  only used the catalogs 
from the Taurus Legacy Program as we do, the study by Luhman et al. 
includes many additional observations of the Taurus region from several General
Observer and Guaranteed Time Observations programs.  
Some well-known Taurus transition disks included in Luhman et. al. 
(e.g., DM Tau and CoKu Tau/4) were not observed as part of the 
Taurus Legacy Program and therefore are not included in our sample. 
The \emph{Spitzer} data for the parent population of our Auriga
targets have not been published at the time of this writing.

\section{Observations}

\subsection{Optical Spectroscopy}

We obtained Echelle spectroscopy   for our entire sample using 
the ESPaDonS Echelle spectrograph on the 3.5-meter Canada-France-Hawaii Telescope (CFHT) 
at Mauna Kea Observatory in Hawaii. The observations were performed in service mode 
over 4 semesters (2009A, 2009B, 2010A, and 2010B).
The spectra were obtained in the standard ``star+sky" mode,
which delivers the complete optical spectra between 3500 \AA \
and 10500 \AA \  at a resolution of 68,000, or 4.4 km/s. 
For each object, we obtained  a set of 3 spectra with exposures times
ranging from 2.5 to 30 minutes each, depending on the brightness
of the target. The data were reduced through the standard CFHT pipeline 
Upena, which is based on the  reduction package 
Libre-ESpRIT\footnote{ \tt http://www.cfht.hawaii.edu/Instruments/Spectroscopy/Espadons/Espadons\_esprit.html}.
In Section~\ref{results}, we use these data to derive the spectral types and accretion rates of our targets.

\subsection{Adaptive Optics Imaging}

High spatial resolution  near-IR observations of our entire sample were obtained in
service mode during the 2009B semester with the 8.1-meter Gemini North telescope  in Mauna Kea
using the Near InfraRed Imager and Spectrometer (NIRI) and the Altair Adaptive Optics (AO) system.
We  used the Natural Guide Star mode for the  targets brighter than R = 15.5 mag,
and the Laser Guide Star mode for the  objects fainter than R=15.5 mag. 
In all cases, the science target was used as the  tip/tilt guide star.
J- and K-band observations were obtained with the f/32 camera (0.022 arcsec/pix, 22$''$ $\times$ 22$''$ field of view).
The brightest objects  required narrow-band filters. 
For each target, we took five  dithered images in  each filter with individual
exposure times ranging from 1 to 20 sec.
The K-band AO images are the most stable and provide the highest Strehl ($\sim$40$\%$), and thus allow the detection of binaries 
with large brightness ratios. The J-band images provide a PSF core with a smaller FWHM 
($\sim$0.05$''$) and are better suitable for the detection of very tight systems. 
The data were reduced with  Image Reduction and Analysis Facility (IRAF)  using the NIREDUCE task in the NIRI package.
In Section~\ref{multi_sec}, we use these AO  imaging data to constraint the 
multiplicity of our targets. 

\subsection{(Sub)millimeter Wavelength Photometry}\label{mm_survey}

The (sub)millimeter wavelength photometry is the most expensive component of our observing program 
in terms of observing time, preventing us from obtaining such data for the entire sample. 
Two of our  targets,  \#2 and 26 (FW Tau),  have already been detected at  (sub)millimeter wavelengths 
(Merin et al. 2010; Andrews \& Williams, 2005), while stringent upper limits  exist for  \#33 
(ZZ Tau; Andrew \& Williams, 2005).  We also found  850 $\mu$m  Submillimeter Array (SMA; Ho et al. 2004) 
archival data  for object \# 9. 
We have observed 26 of the  remaining 37 objects with the SMA. 
We also obtained Common-User Bolometer Array-2 (SCUBA-2)  
data  using the James Clerk Maxwell Telescope (JCMT) for 5 of the  11 targets 
that were not observed with the SMA.
Six objects remain unobserved at (sub)millimeter wavelengths. 
However, since our (sub)millimeter observing programs gave the lowest  priority to objects  with little 24 $\mu$m 
excesses and/or lacking accretion signatures,  these 6 unobserved targets turned out to be either an Asymptotic 
Giant Branch (AGB) stars  contaminating our sample (1 object) or a debris disk candidate 
(5 objects;  see Section~\ref{pms_id}).  
We note that  debris disks are not expected to be detectable at the sensitivity levels of our (sub)millimeter wavelength 
survey,  M$_{dust}$ $\sim$ 6-15 M$_\oplus$ (corresponding to $\sim$2-5  M$_{JUP}$ for primordial disks
with a gas to mass dust ratio of 100). In fact, this dust mass level is an order of magnitude higher 
than that of  the most massive debris disks  observed (Wyatt 2008). 
None of our conclusions are thus  
likely to be affected by the lack of a complete millimeter wavelength  data set.

\subsubsection{Submillimeter Array Observations}

Our SMA observations  were conducted in service mode, during the 2008B,  2009B, and 2010B
Semesters. Virtually all the observations were obtained  in the compact  configuration and with the  
230 GHz/1300 $\mu$m receiver. The only exception was object \#31, which was observed in
the extended configuration using the 345 GHz/850 $\mu$m receiver.
Both the upper and lower sideband data were used, resulting in a total bandwidth of 4\,GHz.

Typical zenith opacities during our 230 GHz observations were $\tau_{225\,{\rm GHz}}$  $\sim$0.15--0.25.
The 345 GHz  extended configuration observations were obtained under significantly better conditions  ($\tau_{225\,{\rm GHz}}$  $\sim$0.07).
For each target, the observations cycled between the target and two gain calibrators (either 3c111 
and 3c84 or 0336+323 and 0449+113),  with 20-30 minutes on target and
 7.5 minutes on each calibrator. 
The raw visibility data were calibrated with the MIR reduction
package\footnote{available at  \tt  http://cfa-www.harvard.edu/$\sim$cqi/mircook.html}.
The passband was flattened using $\sim$1 hour scans of  3c454.3 or  3c279.
The absolute flux scale was determined through   observations of either  
Uranus or Titan and is estimated to be accurate to 15$\%$.
The flux densities of detected sources were measured by fitting a point source model
to the visibility data using the uvmodelfit task in the Common Astronomy Software Applications package
CASA\footnote{available a \tt http://casa.nrao.edu/casa$\_$obtaining.shtml}, while upper limits were derived 
from the rms of the visibility amplitudes. The rms noise of our SMA observations range from $\sim$0.4 to 2.0 mJy per beam. 

We detected, at the 3--$\sigma$ level or better,  4 of the SMA targets: \#5, 29, and 31 (from our 
own data) and  object \# 9 from the archival data. 
Objects \# 2 and 26 from the literature take the number of (sub)millimeter detections to 6. 
The (sub)millimeter wavelength fluxes (and 3-sigma upper limits) for our sample are  listed in Table 2. 
In Section~\ref{disk_mass_sec}, we use the (sub)millimeter wavelength photometry to constrain the masses 
of our transition disks.

\subsubsection{JCMT Observations with SCUBA-2}

The SCUBA-2 data were obtained in service mode on February 28 
and March 1, 2010  during shared-risk observations using only one 
of the 4 arrays per channel of the final SCUBA-2 instrument (Holland et al. 2006).
Each of the 5 objects observed with SCUBA-2 (targets \# 6, 8,  37, 38, and 40) 
were scanned following a daisy pattern for 15 to 25 minutes. During both nights 
the weather was quite good ($\tau_{225}$ = 0.06-0.08);   however, both the 450 and 
850 $\mu$m arrays were significantly noisier than expected during shared-risk observations. 
The strong quasars 3C84 and 3C111 were used as callibrators. 
The data was reduced using the Submm User Reduction Facility 
(SMURF\footnote{http://www.starlink.ac.uk/docs/sun258.htx/sun258.html}).
None of the SCUBA-2 targets  were detected in the final mosaics, where the RMS noise  at 850 $\mu$m
is in the  $\sim$5  to 10 mJ range
The 450 $\mu$m mosaics are much noisier  (RMS $\gtrsim$ 100 mJy) and have little value for our program.  
The  3-$\sigma$ upper limits at 850 $\mu$m for our SCUBA-2 targets are  listed in Table~2.

\section{Results}\label{results}

\subsection{Spectral Types and PMS identification}\label{pms_id}

We estimate the spectral types of our targets by comparing temperature sensitive
features in our echelle spectra to those in templates from stellar libraries. 
We use the libraries presented by Soubiran et al. (1998)  and Montes  (1998). 
The former has a  spectral resolution of 42,000 and covers the  entire 4500 -- 6800  
\AA~spectral range. The latter has a  resolution of 12,000 and covers the  4000 -- 9000  
\AA~spectral range with some gaps in the coverage.  Before performing the comparison, 
we  normalize all the spectra and take the template and  target to a common resolution. 
M-type stars  were  assign  spectral types based  on the strength of the  TiO bands centered 
around 6300, 6700, 7 and 7150 \AA. 
We classify G-K stars based on the ratio of  the V I (at 6199 \AA) to Fe I (6200 \AA) line
(Padgett, 1996) and/or on the strength of the Ca I  ( 6112 A)  and Na I  (5890 and 5896 \AA) 
absorption lines  (Montes et al. 1999;  Wilking et al. 2005).
The F, A,  and B stars were identified and typed by the width of the underlying 
H$\alpha$ absorption line (which is much wider than its emission line) 
and/or  by the  strength of the Paschen 16, 15, 14, and 13 lines.  
The spectral types so derived are listed  in Table 2.
We estimate the typical uncertainties  in our spectral classification to be 1 spectral subclass for M-type 
stars and 2 spectral  subclasses for K and earlier type stars.
Spectral types for most of our targets in Perseus (the IC 348 members)
and Taurus are presented in Luhman et al. (2003), Rebull et al.  (2010),   
and/or Luhman et al. (2010).  We find that most of our spectral types do in fact 
agree within 1 or 2 spectral subclasses with previously published values.

\begin{figure}[th]
\includegraphics[width=5.5in]{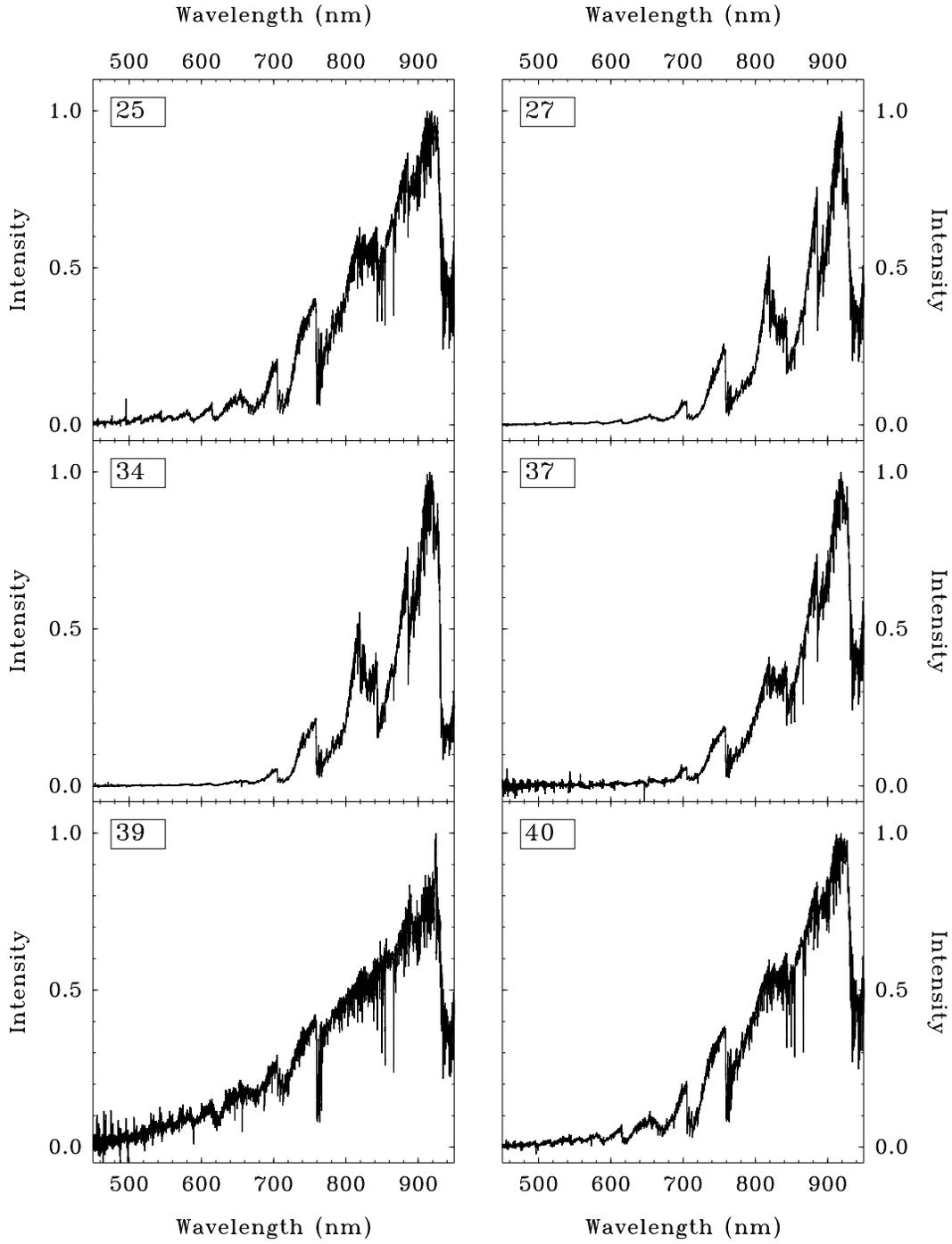}
\caption{
Optical spectra of the 6 AGB stars contaminating our sample}
\label{f:agb_spt}
\end{figure}

Background objects are known to contaminate samples of \emph{Spitzer}-selected 
YSO candidates (Harvey et al. 2007, Oliveira et al. 2009; Papers I and II). 
At low flux levels, background galaxies  are the main source of contamination. 
However, the optical and near-IR flux cuts we have  implemented as part of our sample
 selection criteria are very efficient at removing extragalactic sources
 from the  \emph{Spitzer} catalogs. 
At the bright end of the flux distribution, Asymptotic Giant Branch (AGB) stars 
are the main source of contamination.  
AGB stars are surrounded by shells of dust  and thus have small, but detectable, 
IR excesses.  The extreme  luminosites ($\sim$10$^4$~L$_{\odot}$) of AGB stars 
imply that they can pass our  optical and near-IR flux cuts  even if they are located 
several kpc away. 
AGB stars can easily be identified in our sample as late M-type stars lacking 
both H$\alpha$ emission (from either accretion or chromospheric activity) 
and Li I  6707 \AA~absorption. 
We find 6 such objects in our sample,  all in the Taurus region. 
Their full optical spectra are shown  in Figure ~\ref{f:agb_spt}. 
As seen in  Figure~\ref{f:sample_sel},  the contamination of AGB stars is particularly high in the 
[3.6]-[24]  colors range between 1.5 and  2.5  (five of the six AGB stars fall in this narrow range).

\begin{figure}[th]
\includegraphics[width=6.0in]{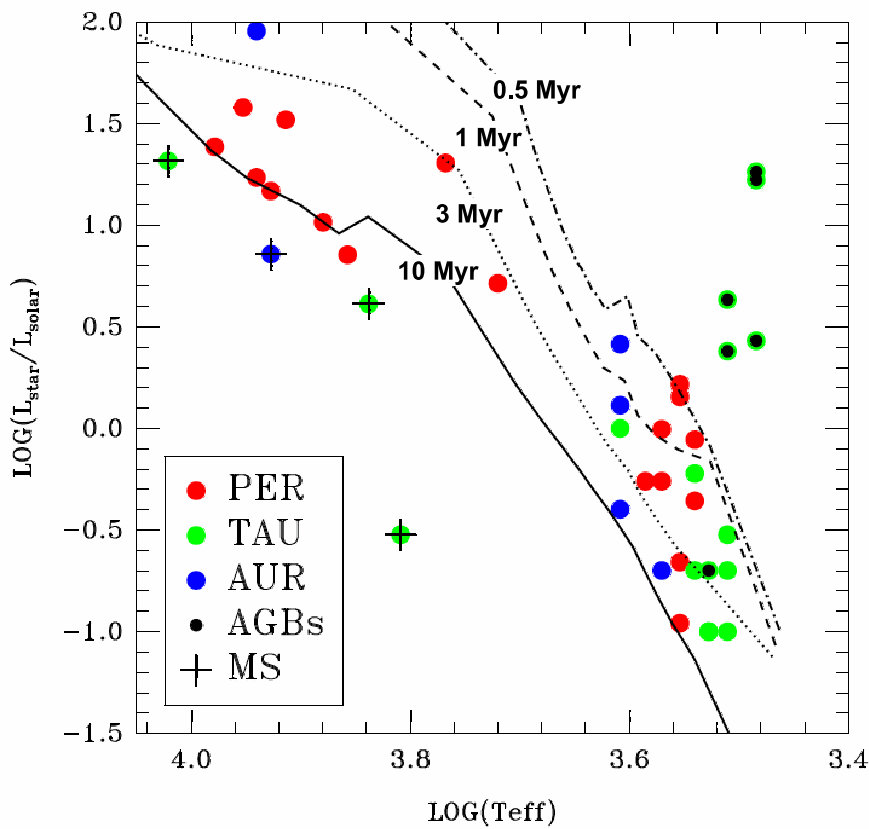}
\caption{
The location of all our targets in the H$-$R diagram diagram. The 0.5, 1, 3, and 10 Myr
isochrones from  the Siess et al. (2000) models are also shown.
Five of the six AGB stars are clearly overluminous (targets \# 25, 27, 34, 37, and 40).  
Three stars in Taurus (\# 35, 36,  and 38) and one in Auriga (\# 30) are significantly underluminous 
and are likely to be background MS stars. One target in Perseus  (\# 11) is also slightly underluminous, 
but it has been classified as a member of the IC 348 cluster based on  proper motion measurements 
(Luhman et al. 2003). 
}
\label{f:HRD}
\end{figure}

The Li I  6707 \AA~absorption  line is a very good indicator of stellar youth 
in mid-K to M-type stars because Li is burned very efficiently in the convective 
interiors of low-mass stars and is depleted soon after these objects arrive on the 
main-sequence (Cargile et al. 2010); however, it is not an age discriminant for early type stars where 
the depletion timescales can approach the main-sequence ages. 
To establish whether our early type targets are in fact consistent with pre-main-sequence
(PMS) stars associated  with the Perseus, Taurus, and Auriga molecular clouds, 
we place them in the Hertzsprung--Russell (H-R) diagram (see Figure~\ref{f:HRD}) 
and compare their position against the theoretical isochrones from Siess et al. (2000). 
We adopted the bolometric corrections and temperature scale
from Kenyon $\&$ Hartmann (1995) and  the following  distances:  
320 pc for Perseus  (Herbig 1998),  140 pc for  Taurus 
(Kenyon, G{\'o}mez \& Whitney,  2008), and 300 pc for Auriga (Heiderman et al. 2010).  
We corrected for extinction using  A$_{J}$ =1.53 $\times$ E(J-K$_{S}$), 
where E(J-K${_S}$) is the observed color excess  with  respect to the expected 
photospheric color for the given spectral type (also from Kenyon $\&$ Hartmann 
1995).
We find that 5 of our early type (F5 to B9) targets fall below the 10 Myr isochrone (\# 11, 30, 35, 36, and 38). 
All of them have very low disk luminosities and could be background MS stars with debris disks. MS 
stars in this temperature range are 2.5 to 100 times  more luminous than the Sun and can be seen 
at large distances.  Their large luminosities also imply that their debris disks are more likely to 
be detectable at \emph{Spitzer} wavelengths because the fractional disk luminosity needed to produce 
a detectable 24 $\mu$m excess is much lower for BAF-type stars than it is for lower luminosity objects
(Cieza et al. 2008a).   However,  object \# 11 in Perseus has been identified as a member of IC 348
cluster (Luhman et al. 2003) based on proper motion measurements. Given the uncertainties
associated with placing objects in the H-R diagram, we consider object \# 11 to be 
a PMS star. In Section~\ref{classification}, we classify the other four under-luminous 
objects as MS debris disk candidates.

The H-R diagram also serves as an independent check for our sample of  late-type targets.
As seen in Figure~\ref{f:HRD}, all the K and M-type objects fall within the 0.5 and 10 Myr isochrones,
except for 5 of the 6 AGB stars which are clearly too over-luminous to be low-mass objects
at the distance of the Taurus molecular cloud. 
Setting aside 6 AGB stars (targets \# 25, 27, 34, 37, 39 and 40), and 4 likely MS stars (sources \#30, 35, 36, and 38) 
we are left with  31 objects that  constitute our sample of transition disks around bona fide PMS stars. 

Unlike Papers I and II,  where we found that virtually all the  objects were K and M-type 
stars,  here we find that  
$\sim$30$\%$ of the bona fide transition objects  (9/31)
are BAFG-type stars.    
This difference is likely to be due to a combination of several effects.
First, the Initial Mass Function (IMF) is not the same in all the  clouds. 
Lupus and Ophiuchus contain a larger proportion of very  low-mass stars  (i.e., late M-type objects) than Taurus 
(Hughes et al. 1994; Erickson et al. 2011). 
Second, many low-mass objects at the distance of Perseus and Auriga (300-320~pc)
are too faint to meet all of our sample selection criteria.
Finally, there are significant differences in the sizes of the PMS populations.
Since there are $\sim$3 more YSO candidates in Perseus than in Lupus,  Perseus should 
contain a larger number of high mass stars for a given IMF. 

\subsection{H$\alpha$ Profiles and Accretion Rates}\label{Acc}

Most young low-mass stars show H$\alpha$ emission, either from chromospheric activity or 
accretion.  Non-accreting objects show narrow ($\lesssim$ 200 km/s, measured at 10 $\%$ of the peak intensity) 
and  symmetric line profiles of chromspheric origin, while accreting objects present broad ($\gtrsim$ 270 km/s) 
and asymmetric profiles produced by large-velocity magnetospheric  accretion columns.
In order to measure the velocity width of the H$\alpha$ emission line, we first subtract 
the continuum by fitting a line to the spectrum in the velocity intervals
 $-$600 to $-$400 km/s and 400 to 600 km/s  centered at the H$\alpha$ location. 
The peak of the continuum-subtracted  spectrum is then normalized to 1,  and the velocity 
width, $\Delta V$,  is measured at  10 $\%$ of the peak intensity. 
The  velocity dispersion  so derived for  the H$\alpha$ emission 
lines of our  transition disks are listed in Table 2.
We estimate the typical uncertainty in $\Delta V$ to be of the order of 10$\%$. 
However, as discussed  below, the uncertainty in the corresponding accretion rate is expected to be 
much larger. 

\begin{figure}
\includegraphics[width=5.5in]{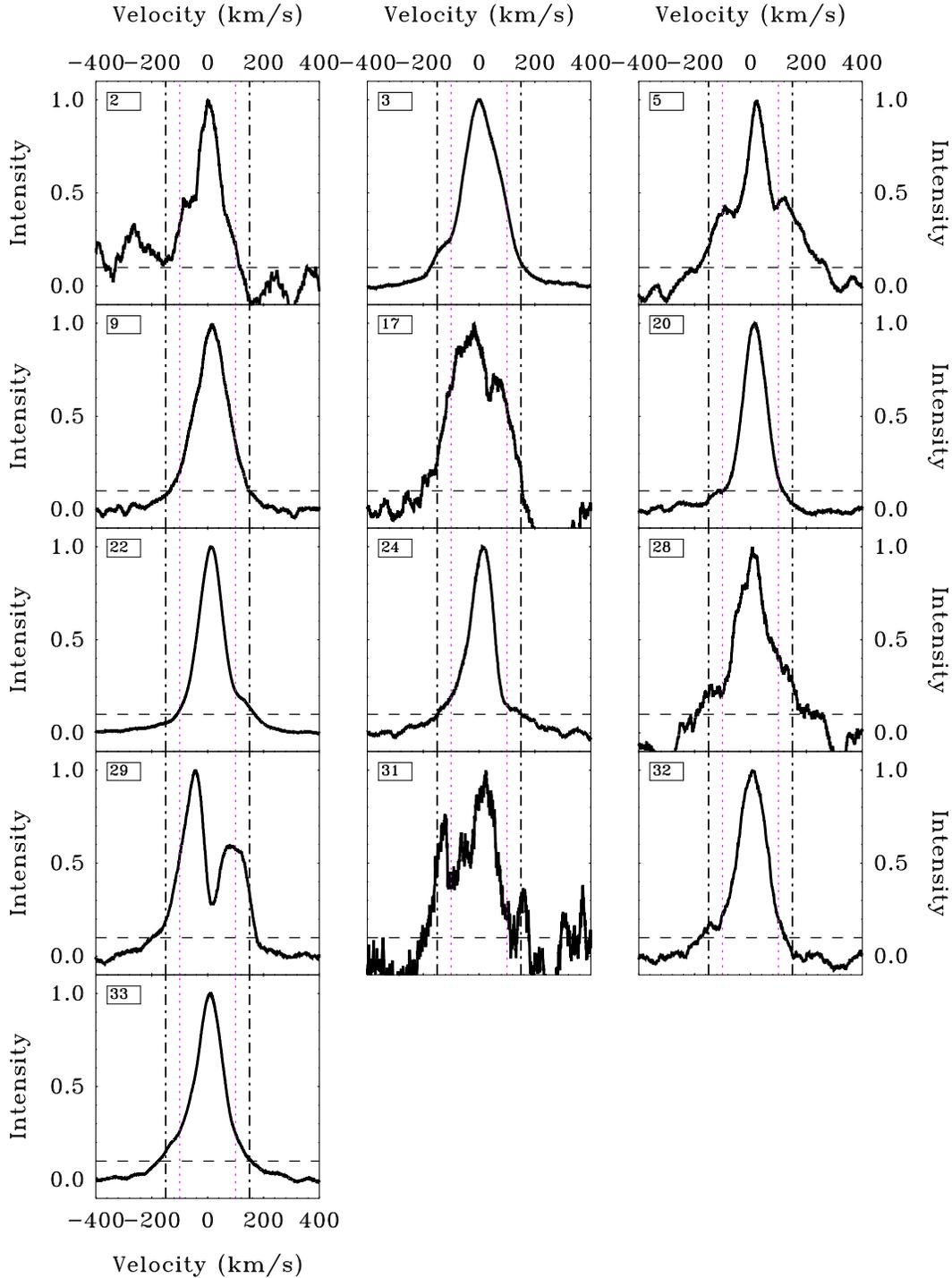}
\caption{The H$\alpha$ velocity profiles of the 13  accreting 
objects in our sample.  The dashed line indicates the 10$\%$ peak intensity,
where $\Delta$V is measured. The intervals delimited  by the dotted and dashed-dotted lines 
correspond to $\Delta$V~=~200 and $\Delta$V~= ~300 km/s, respectively. All  accreting objects 
have $\Delta$V $>$ 200 km/s.
}
\label{f:prof_acc}
\end{figure}

The boundary between accreting and non-accreting objects has been empirically placed by
different studies at $\Delta V$ between 200 km/s (Jayawardana et al. 2003) and 270 km/s 
(White $\&$ Basri, 2003). 
Since only two objects, sources \#  20  and \#23  have  $\Delta V$  in the 200-270 km/s range,  
most accreting and non-accreting objects are clearly separated in our sample. 
Source \# 20 is a M6 star with a rising mid-IR SED, a 70~$\mu$m detection, and $\Delta V$ $\sim$ 210 km/s. 
Very low-mass stars tend to have narrower  H$\alpha$ lines than higher mass objects 
because of their  lower accretion rates (Nata et al. 2004) and their lower gravitational potentials 
(Muzerolle et al. 2003).   Given all the available data, we classify  target \# 20 as an
accreting object, 
but warn the reader that its accreting nature is less certain than that of the 
rest of the objects classified as CTTSs.  
Similarly, source \# 23 has $\Delta V$ $\sim$200 km/s and we consider it to be non-accreting because of its
very low fractional disk luminosity (L$_D$/L$_{*}$ $<$ 10$^{-3}$, see \S~\ref{sed_mor}), and 
it is most likely an optically thin debris disks; however, its non-accreting nature is not certain. 
The continuum-subtracted H$\alpha$ profiles for all  the 13 accreting transition disks are shown in 
Figure~\ref{f:prof_acc}. 

\begin{figure}[th]
\includegraphics[width=5.5in]{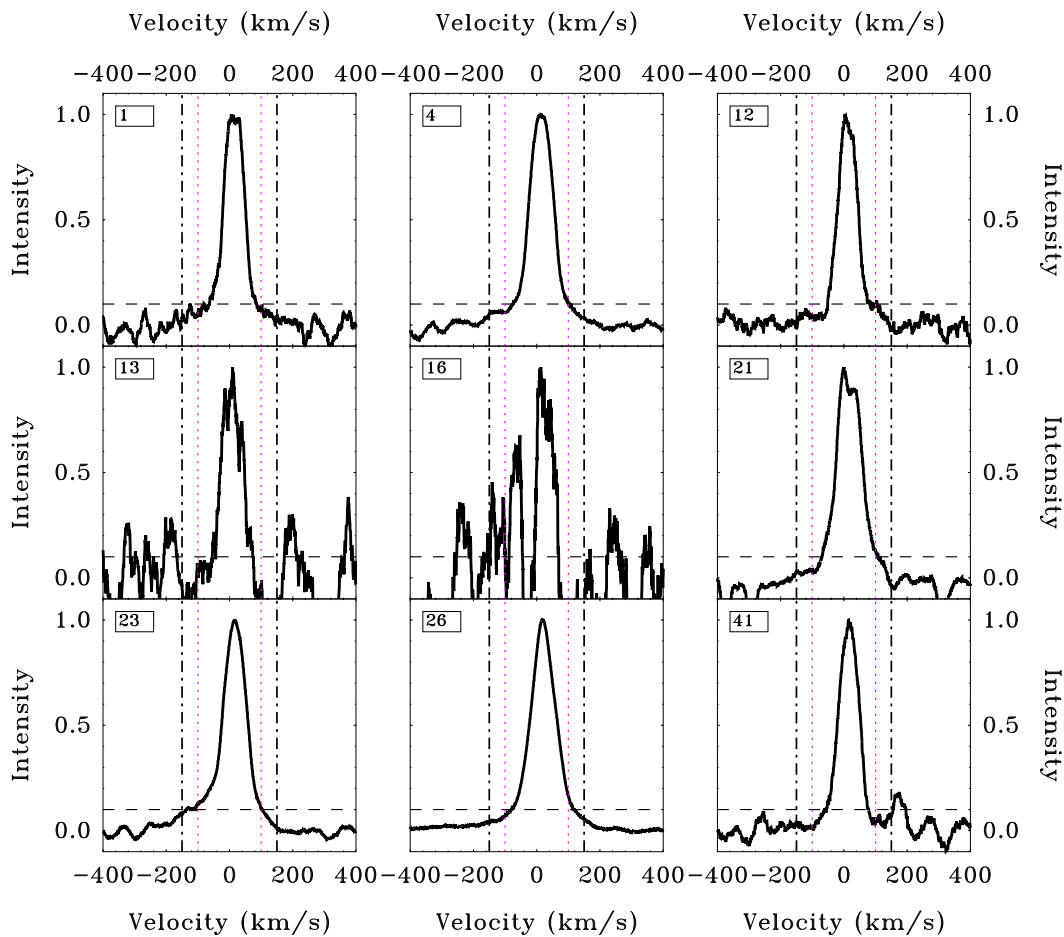}
\caption{
The H$\alpha$ velocity profiles of the 9 non-accreting 
objects in our sample where H$\alpha$ was detected in emission.
They are all K and M-type stars.
The dashed line indicates the 10$\%$ peak intensity,
where $\Delta$V is measured.
The intervals delimited  by the dotted and dashed-dotted lines 
correspond to $\Delta$V~=~200 and $\Delta$V ~= ~300 km/s, respectively. 
Non-accreting objects  show symmetric and  narrow  
($\Delta$V $\lesssim$ 200 km/s)
H$\alpha$ emission, consistent with chromospheric
activity.  
}
\label{f:prof_non_acc}
\end{figure}

There are 9 K and M-type stars in our sample with narrow H$\alpha$ emission
consistent with chromospheric activity. The continuum-subtracted H$\alpha$ profiles
of these objects are shown in Figure~\ref{f:prof_non_acc}. 
The spectra of targets \# 13 and 16 have lower  signal to noise ratios than
the rest of the spectra in the figure, which makes it difficult to measure 
accurate $\Delta V$ values. 
%
%
Target \#~13  is an M2 star with a small H$\alpha$ 
equivalent widths  (3.5 \AA; Luhman et al. 2003)  and  object \#~16 has 
L$_D$/L$_{*}$ $<$ 10$^{-3}$ and it is another likely  debris disk.
For  disk classification purposes, we consider both objects to be non-accreting. 
The 13 BAFG stars in our sample show the characteristic photospheric H$\alpha$ line 
in absorption, with little  or no evidence for significant superimposed H$\alpha$ emission. 
As discussed in the previous section, 4 of them are likely to be background MS 
star (objects \# 30, 35, 36, and 38). The other 9 objects are consistent with
non-accreting PMS stars and are bona fide targets.

For accreting objects, the velocity dispersion of the H$\alpha$ line
correlates well with accretion rates derived from models of the 
magnetospheric accretion process. We therefore estimate the accretion rates of our targets
from the width of the H$\alpha$ line measured at 10$\%$   
of its peak intensity, adopting the relation given by Natta et al.  (2004):

\begin{equation}
Log (M_{acc}(M_{\odot}/yr)) = -12.89(\pm0.3) + 9.7(\pm0.7)\times10^{-3} \Delta V (km/s)
\end{equation} 

This relation is valid for 600 km/s $>$ $\Delta V$ $>$  200 km/s  (corresponding to  
10$^{-7}$ M$_{\odot}$/yr $>$ M$_{acc}$  10$^{-11}$ M$_{\odot}$/yr) and can be applied 
to objects with a range of stellar (and sub-stellar)  masses.  
As discussed by Muzerolle et al. (2003) and Sicilia-Aguilar et al. (2006b), 
the broadening of the H${\alpha}$ line is  1 to 2 orders of 
magnitude more sensitive to \emph{low} accretion rates than other accretion 
indicators such as U-band excess and continuum veiling measurements, and is thus
particularly useful to distinguish weakly accreting from non-accreting 
objects. However, the 10$\%$
width measurements are also dependent on the line profile, rendering the
10$\%$ H$\alpha$ velocity width a relatively poor \emph{quantitative} 
accretion indicator (Nguyen et al.  2009). 

For the objects we consider to be non-accreting, we adopt a mass accretion  upper limit of 10$^{-11}$M$_{\odot}$/yr, 
corresponding to $\Delta V$ = 200 km/s,  although the detectability of accretion is both a function of 
spectral type and data quality. 
The  so derived  accretion rates (and upper limits) for our sample of transition disks are listed in Table 3.
Given the large uncertainties associated with Equation 2 and the intrinsic 
variability of accretion in PMS stars, these accretion rates should be 
considered to be order-of-magnitude estimates.

\subsection{Disk Masses}\label{disk_mass_sec}

Disk masses obtained from modeling  the  IR and (sub)millimeter SEDs of circumstellar 
disks are well described by a simple linear  relation between (sub)millimeter flux and disk mass 
(Andrews $\&$ Williams  2005; 2007). 
Following Papers I and II, we adopt the linear relations presented by 
Cieza et al. (2008b): 
 
\begin{equation}
 M_{DISK}=1.7\times10^{-1}  [(\frac{F_\nu(1.3mm)}{mJy})\times(\frac{d}{140  pc})^2] M_{JUP} 
 \label{eq_mass}
\end{equation}

\begin{equation}
 M_{DISK}=8.0\times10^{-2}  [(\frac{F_\nu(0.85mm)}{mJy})\times(\frac{d}{140  pc})^2] M_{JUP} 
 \label{eq_mass}
\end{equation}

These relations come from the ratios of model-derived  disk masses  to observed (sub)millimeter fluxes  
presented by  Andrews $\&$ Williams (2005) for 33 Taurus stars.
Based on the standard deviation  in the ratios of the model-derived  masses to observed (sub)millimeter fluxes, 
the above relation gives disk masses  that are within a factor of $\sim$2 of model-derived values; 
nevertheless,  much larger \emph{systematic} errors can not be ruled out (Andrews $\&$ Williams, 2007). 
In particular, the models from Andrews $\&$ Williams, 2005; 2007)
assume an opacity as a function of frequency of the form $K_{\nu}$ 
$\propto$ $\nu$  and a normalization of $K_0$ =  0.1 gr/cm$^2$ at 1000 GHz. 
This opacity implicitly assumes a gas to dust mass ratio of 100. 
Both the opacity function and the gas to dust mass ratio are highly
uncertain and expected to change due to disk evolution processes 
such as grain growth and photoevaporation.  Detailed modeling and
additional observational  constraints on the grain size distributions (e.g., from sub/millimeter wavelength
slopes) and the gas content (e.g., from CO, [O~I], and/or [C~II] observations)  will be needed to derive 
more accurate disk masses for each individual transition disk.

The disk masses (and 3-$\sigma$ upper limits) for our sample are listed in Table 3 and were 
derived adopting the distances to the clouds from Section~\ref{pms_id}.
The vast majority of our transition objects have estimated disk masses lower than $\sim$1--3 M$_{JUP}$.
However,  5 of  them have disk masses typical  of CTTSs  ($\sim$3--15 M$_{JUP}$).

\subsection{Stellar companions}\label{multi_sec}

\begin{figure}[t]
\includegraphics[width=6.5in]{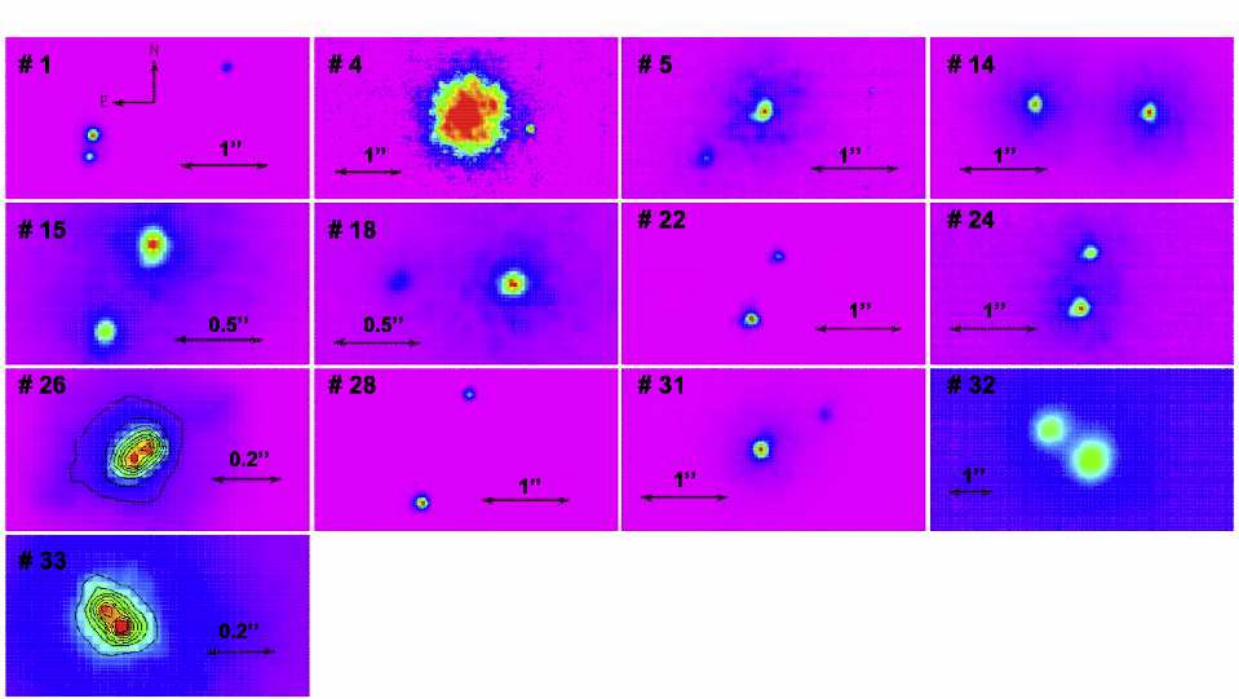}
\caption{The AO images of the 13 multiple systems that have been detected with our Gemini observations.
Target \#1 is a triple system.  
The images shown correspond to the K-band, except for objects \# 26, and 33, for which
the J-band images are shown because they have a somewhat higher spatial resolution. 
}
\label{f:multi}
\end{figure}

From the visual inspection  of our Gemini-AO images, we identify 13 multiple systems:
targets \# 1, 4, 5, 14, 15, 18, 22, 24, 26, 28, 31, 32, and 33 (see Figure~\ref{f:multi}).  
The separation and flux ratios of these systems range from 
0.05$''$ to 1.7$''$ and 1.0 to 14, respectively. 
Object \# 1 is a triple. 
Most  systems were fully resolved in both our J- and K$_{S}$-band images, which 
have typical FWHM values of 0.06$''$-0.08$''$. The tightest systems, targets \# 26 and 
33, were only fully resolved  by the J-band images (FWHM $\sim$ 0.05$''$). 

In Perseus, targets \#~14 and 15 are previously known binaries (Duchene et al.  1999), while targets \#~1, 4, 5, and 18 are newly identified multiple systems. 
Similarly, objects \#~24, 26, and 33 are known Taurus binaries (Koponacky et al. 2007; Simon et al. 1995), while targets \#~22 is a newly
discovered Taurus binary system.
All 3 multiples in Auriga (objects \#~28, 31, and 32)  are newly identified systems. 
For target \#~32, the AO system did not lock correctly in either of the two components  of the binary, resulting in
a very poor AO correction and a much uncertain determination of the flux ratio.

For the apparently single stars, we estimated the detection limits at 0.1 and 0.2$''$  separations from the 5-$\sigma$ noise 
of PSF-subtracted images. 
Since no PSF standards were observed in our program, 
we subtract a PSF constructed by azimuthally smoothing the  image 
of the target itself,  as follows. For each pixel in the image, the separation from the target's centroid is calculated, 
with sub-pixel accuracy. The median intensity at that separation, but within an arc  of  30 pixels in length, 
is then subtracted from the target pixel.  Thus, any large scale, radially symmetric  structures are removed.
The separations, positions angles, and flux ratios of the multiple systems  in our sample 
are shown in Table~2.  The flux ratio limits for unseen companions at 0.1 and 0.2$''$  separations, 
obtained as described above, are also listed for the targets that appear to be single.

\subsection{SED morphologies and fractional disk luminosities}\label{sed_mor}

In addition to the disk mass, accretion rate, and multiplicity,  the SED morphology and fractional disk
luminosity of a transition disk can provide important clues on  the nature of the object.
Following Papers I and II,  we quantify the SED morphologies and fractional disk luminosities of 
our transition objects and use these quantities as part  of our disk classification  scheme (see Section~\ref{classification}).
We quantify  the  SED  ``shape" of our targets adopting the two-parameters 
introduced by Cieza et al. (2007): $\lambda_{turn-off}$, which is  the longest wavelength at which the 
observed flux is dominated by the stellar photosphere,  and $\alpha_{excess}$, the slope 
of the IR excess,  computed as  dlog($\lambda$F)/dlog($\lambda$) between 
$\lambda_{turn-off}$ and 24 $\mu$m.  
To calculate $\lambda_{turn-off}$, we compare the extinction-corrected SED to NextGen Models
(Hauschildt et al. 1999) normalized to the J-band and choose $\lambda_{turn-off}$  as the longest 
wavelength at which the stellar photosphere contributes over 50$\%$ 
of the total flux. The uncertainty of  $\lambda_{turn-off}$  is roughly one SED point.
The $\lambda_{turn-off}$ and  $\alpha_{excess}$ values for our entire sample are listed in Table 3. 
The $\lambda_{turn-off}$ and  $\alpha_{excess}$ parameters are dependent on 
the SED sampling and are affected by IR variability (Espaillat et al. 2011); however, they  
provide first-order information on the structure of the disk.  
For a given stellar luminosity, 
the $\lambda_{turn-off}$ value correlates with the size of the inner hole as it depends on 
the temperature of the dust closest to the star.  
It is clear however, that a given $\lambda_{turn-off}$  value implies a much larger
inner hole for a disk around an A-type star than for one around a M-type star. 
Similarly,  $\alpha_{excess}$ correlates well with the sharpness of the opacity hole.
On the one hand, sharp inner holes result in positive $\alpha_{excess}$ values,
which are typical of  classical transition disks (Muzerolle et al. 2010) and  cold disks 
(Brown et al. 2007; Merin et al. 2010) with rising mid-IR SEDs.
On  the other hand,  more radially continuous disks that have undergone significant grain 
growth and dust settling  show  the  negative $\alpha_{excess}$ values (Dullemond $\&$ Dominik, 2004) that is
characteristic of anemic (Lada et al. 2006) and homologously depleted  (Currie et la. 2009) transition disks 
with falling mid-IR SEDs.
See Espaillat et al. (2012) for a recent discussion on the link between SED morphology and physical 
properties of the disk.

The fractional disk luminosity, the ratio of the disk luminosity to the
stellar luminosity (L$_{D}$/L$_{*}$), is another important quantity that
relates to the evolutionary state of a disk.  Typical primordial 
disks around CTTSs have L$_{D}$/L$_{*}$ $\sim$ 0.1 as they have 
optically thick disks that intercept (and reemit in the IR) $\sim$10$\%$ of the 
stellar radiation.  In contrast, debris disks show  L$_{D}$/L$_{*}$ 
values $\lesssim$ 10$^{-3}$  because they have optically thin disks that intercept 
and reprocess $\sim$10$^{-5}$--10$^{-3}$ of the star's light  (Bryden et al. 2006). 

We estimate L$_{D}$/L$_{*}$  for our sample of  disks
by integrating over frequency the flux contribution of the disk and  the star to the observed 
SED (see Paper I for calculation details). 
The near and mid-IR luminosities of our disks are well constrained because their SEDs are  
relatively well sampled at  these wavelengths. However, their far-IR  luminosities remain
more poorly constrained.  Only 5 of our 41  targets have (5-$\sigma$ or better)  detections  
at 70 $\mu$m listed in the \emph{C2D}, \emph{Goulds Belt},  and \emph{Taurus}  catalogs.  
For the rest of the objects, we have obtained 5-$\sigma$ upper limits as in Paper I 
(from the noise of the 70 $\mu$m images at the location of the targets)  in order to fill the gap in their
SEDs between 24 $\mu$m and the millimeter. 

The LOG(L$_D$/L$_*$) values for our transition disk sample, ranging from $-$1.7 to $-$4.9 
are listed in Table 3.  
The LOG(L$_D$/L$_*$) values are highly dependent on SED sampling and should 
be considered order of magnitude estimates. 
As most of the luminosity of a disk extending inward to the  dust sublimation temperature is emitted in the near-IR, L$_D$/L$_{*}$
 is a very strong function of $\lambda_{turn-off}$: the shorter the $\lambda_{turn-off}$ wavelength, the higher the 
fractional disk luminosity.  
For objects with $\lambda_{turn-off}$ $<$ 8.0 $\mu$m,  the 70 $\mu$m flux represent only a minor contribution to the total 
disk luminosity. 
On the other hand, objects with with  $\lambda_{turn-off}$ $=$ 8.0 $\mu$m (i.e., objects where the IR excess only 
becomes significant  at 24 $\mu$m) have much lower L$_D$/L$_*$ values, and the 70 $\mu$m emission 
becomes a much larger fraction of the total disk luminosity.  As a result, the L$_D$/L$_*$ values of objects 
with $\lambda_{turn-off}$ $=$ 8.0 $\mu$m and no 70 $\mu$m detections should be considered upper limits.

\section{Discussion}

 \subsection{Disk Classification}\label{classification}
 
 One of the main results from Papers I and II has been the very large range of disk
 properties (accreation rates, disks masses, SED morphologies, and L$_D$/L$_*$ values) 
 exhibited by our targets. This diversity points toward different evolutionary stages and/or
 different physical processes driving the evolution of each disk. 
 In particular, the  wealth of information discussed in the previous sections
 allow us  to place each target in our transition sample into the following categories:
  grain-growth dominated disks, giant planet-forming disks,  
  photoevaporating disks, debris disks, and circumbinary disks.
 
 \subsubsection{Grain-growth dominated disks:}
 
\begin{figure}[t]
\includegraphics[width=5.0in]{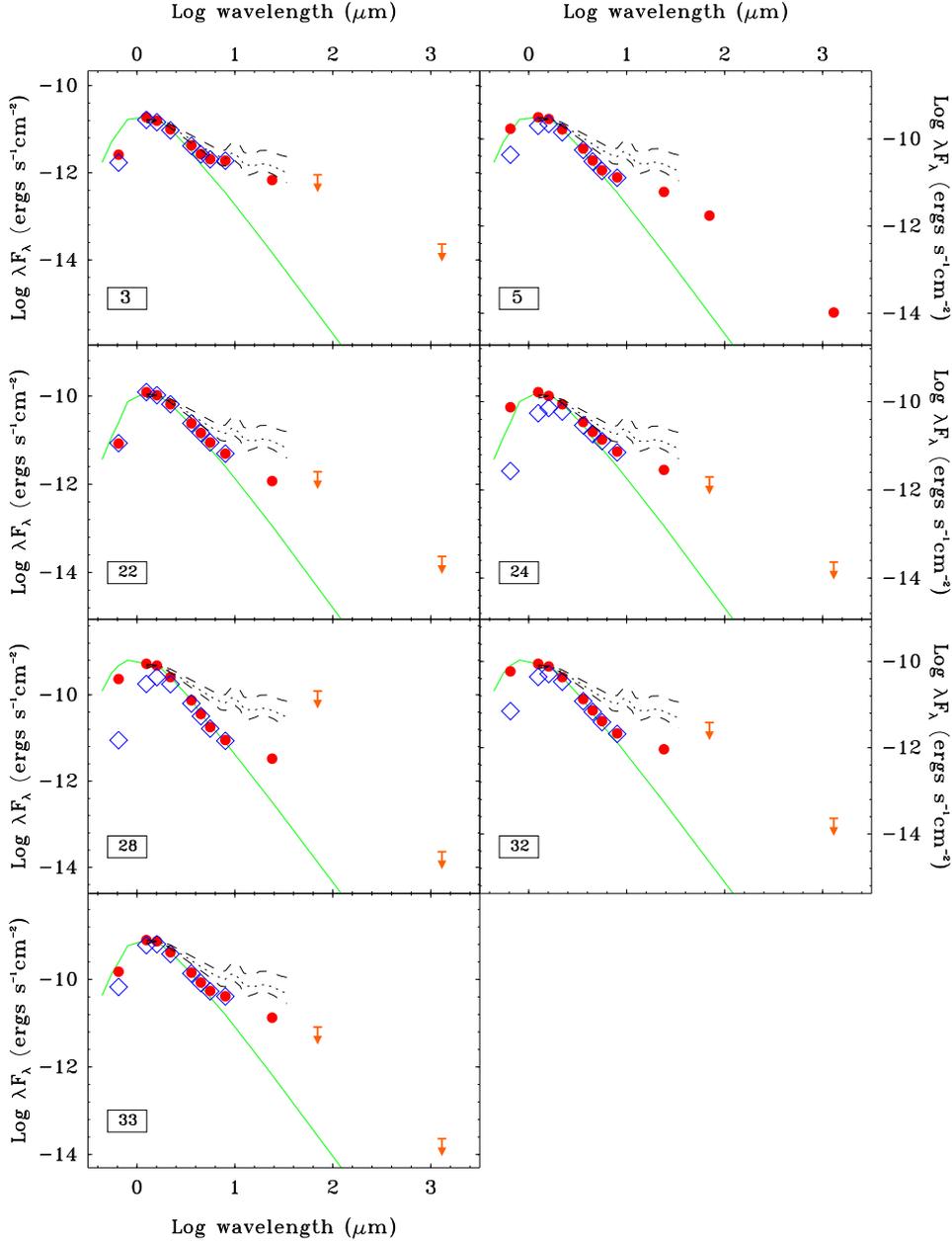}
\caption{The SEDs of the  7 grain-growth dominated disk candidates.
The  filled circles are detections while the arrows represent 3-$\sigma$ limits. The 
open squares correspond to the observed optical and near-IR fluxes before 
being corrected for extinction using the A$_J$ values listed in Table 3 (calculated as described 
in Section~\ref{pms_id})  and the 
extinction curve provided by the Asiago database of photometric systems  
(Fiorucci $\&$ Munari  2003). 
The solid green lines represent the stellar photosphere normalized to the extinction-corrected J-band.  
The dotted lines correspond to the median mid-IR SED of K5--M2 CTTSs calculated by Furlan et al. (2006). 
The dashed lines are the quartiles.  
}
\label{f:sed_GGD}
\end{figure}

Accreting objects with  $\alpha_{excess}$ $<$  0 (falling mid-IR SEDs)
are most readily explained as primordial disks that have 
undergone significant  grain growth and dust settling.  
The SEDs of  the 7 transition disks in this category are shown in  Figure~\ref{f:sed_GGD}.
Circumstellar disks are initially very flared and intersect a significant fraction of  the stellar 
radiation, most of which is reprocessed and reemitted at IR wavelengths. As dust grains 
coagulate and  grow in the disk, they fall toward the mid-plane, where the surface density 
is higher and they can grow at a higher rate and settle even deeper into the disk.
As a result, the disk becomes geometrically flatter with time,  which reduces the 
fraction of stellar radiation intercepted by the disk and steepens the slope of the SED at
mid-IR  wavelengths (D'Alessio et al. 2006; Dullemond et al. 2007).  
Grain growth is expected to proceed faster in the inner disk, where the
surface densities are higher and the dynamical timescales are shorter.
Thus, the depletion of micron-sized grains in the inner disk through grain growth 
also  contributes to the low levels of near and mid-IR excesses seen in this type 
of objects (Dullemond $\&$ Domink, 2005). 

Based on their SEDs, grain-growth dominated disks could also be classified 
as anemic (Lada et al. 2006) or homologously depleted transition disks 
(Currie et la. 2009). They show a large diversity  of accretion rates  and disk masses, 
which in this paper range from  10$^{-10.3}$ to 10$^{-8.5}$  M$_{\odot}$ yr$^{-1}$
and $<$ 0.5  to 5.3 M$_{JUP}$ (but note DoAr 25 in Paper I, with a disk mass of
$\sim$40 M$_{JUP}$ and an accretion rate of  10$^{-7.2}$  M$_{\odot}$ yr$^{-1}$).
Establishing whether the decrement of IR excess is mostly due to a reduction of 
the dust opacity, a reduction of the surface density in the disk, or purely 
geometrical effects requires  resolved observations and/or details modeling
of individual objects. 
In some cases, the weakness of the IR excess might be attributed to the very 
low luminosity of the central star (see Section 5.2.4)."

\subsubsection{Giant planet-forming disks}\label{PFD_def}  

\begin{figure}[t]
\includegraphics[width=5.5in]{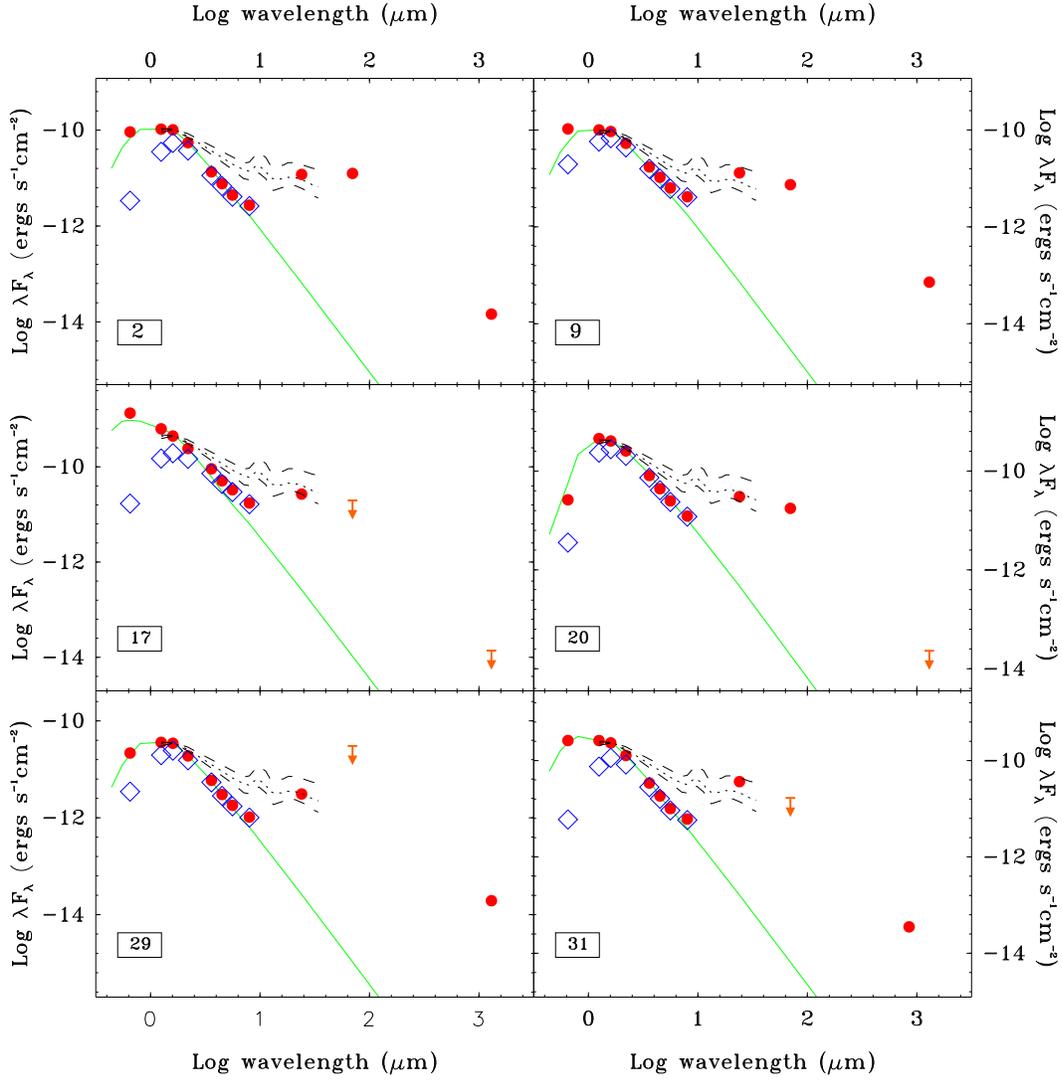}
\caption{The SEDs of the 6  giant planet-forming disk candidates. 
The symbols are the same as in Figure~\ref{f:sed_GGD}. 
}
\label{f:sed_PFD}
\end{figure}

Accreting transition disks  with sharp inner hole (i.e., $\alpha_{excess}$ $\gtrsim$ 0; rising mid-IR SEDs) 
are currently  considered the most likely sites for ongoing planet formation 
(Najita et al. 2007;  Merin et al. 2010; Williams \& Cieza, 2011). 
The recent identification of planet candidates within the inner holes of the  T Cha (Hu\'elamo et al. 2011)
and LkCa 15  disks (Kraus $\&$ Ireland,  2011) strongly supports this interpretation.  
We find that 6 objects fall in this category (see Figure~\ref{f:sed_PFD}). 
The presence of accretion unambiguously identify them
has gas-rich primordial disks. 
The connection between a  rising mid-IR SEDs  and the presence of a sharp inner hole 
(a large change in surface density  over an small radial distance) is well established 
through both SED modeling (Brown et al. 2007; Calvet et al. 2002; 2005)  and direct submillimeter 
imaging (Brown et al. 2009; Hughes et al. 2007; 2009).
While in addition to planet formation, other processes such as grain growth, photoeveporation, and   
dynamical clearing due to \emph{stellar} companions have been proposed to
explain this type of objects,  these alternative explanations face serious theoretical and/or 
observational challenges. 
In particular, grain growth efficiency is expected to be an smooth function   
of radius, which is inconsistent with the abrupt change in opacity
inferred for transition disk with steeply rising mid-IR SEDs. 
Whether photoevaporation can account for the inner holes
of accreting objects depends on the photoevaporation rates.
As will be discussed in Section~\ref{photo_constrain},  there is
strong evidence \emph{against} photoevaporation rates being large enough 
to explain the inner holes of accreting transition disks.
We thus conclude that the presence of accretion makes the photoevaporation scenario much less likely.

Dynamical clearing by (sub)stellar companions (Lubow \& D'Angelo, 2006; Zhu et al. 2011)
therefore remains the most likely explanation for accreting transition disks with rising mid-IR SEDs.
Recent hydrodynamical simulations of  multiple planets embedded within the disk also help reconciling 
the lack of near-IR excess with the levels of accretion seen in  transition disks in this category.
The optically thick but
physically narrow tidal tails predicted by these simulations can transport significant  amounts 
of material  without  over-predicting the IR excesses observed (Dodson-Robinson \& Salyk, 2011).  
While our AO observations cannot rule out the presence of stellar companions to our targets 
inward of ($\sim$15-30 AU), Near-IR interferometry (Pott et al. 2010)  and aperture masking observations
(Kraus et al. 2009)
conclusively exclude \emph{stellar} companions as a possible cause of the inner holes
in several similar objects. 
All things considered, we conclude that the dynamical clearing by one or multiple planets is
the most likely explanation for the properties of the 6 objects shown in Figure~\ref{f:sed_PFD}.

\subsubsection{Photoevaporating and debris disks}

We find that virtually all non-accreting objects remain undetected at millimeter wavelengths, implying 
dust masses  below 6--10 M$_\oplus$ (corresponding to  disk masses $\lesssim$ 2--3 M$_{JUP}$ for a gas-rich disk
with a gas to dust mass ratio of 100).  Target \# 26 (FW Tau) with an estimated  disk mass of 
0.4 M$_{JUP}$ is the only exception.
All photoevaporation models predict the formation of 
a gap in the disk and the subsequent draining of the inner disk once most of the outer disk mass has been
depleted and the accretion rate falls below the photoevaporation rate. 
Once this inner hole has formed,  the high-energy photons can reach the inner edge of the disk unimpeded, 
and the disk quickly dissipates from the inside out. 
After the circumstellar gas photoevaporates,  the small grains are removed by radiation pressure and 
Poynting-Robertson  drag, leaving behind a gas-poor disk with large grains, planetesimals and/or planets 
(i.e., a debris disk).  
In this context, the most fundamental difference between a photoevaporating disk and 
a young debris disk is the presence of primordial gas at large radii: 
if the inner hole is due to photoevaporation, the outer disk should remain gas-rich beyond 
the photoevaporation front.  
Since we lack information on the gas content in  the outer disks of  our targets, we adopt a  less
direct criterion based on the fractional disk luminosity, L$_D$/L$_*$, to  tentatively distinguish 
primordial  photoevaporating disks from debris disks.

\begin{figure}[t]
\includegraphics[width=5.0in]{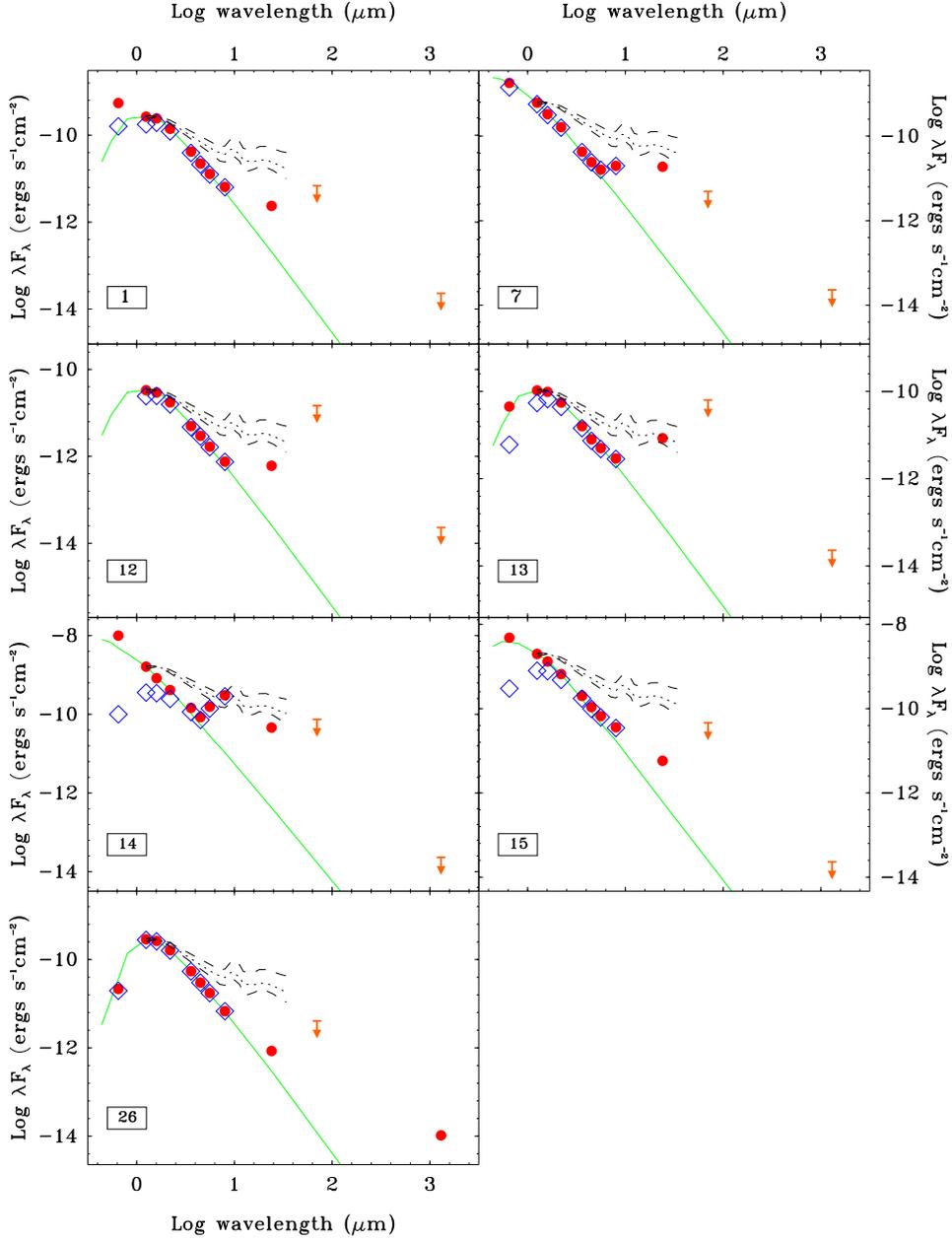}
\caption{The SEDs of the 7 photoevaporating disk candidates.
The symbols are the same as in Figure~\ref{f:sed_GGD}.
}
\label{f:sed_photo}
\end{figure}

\begin{figure}[htp]
\includegraphics[width=6.5in]{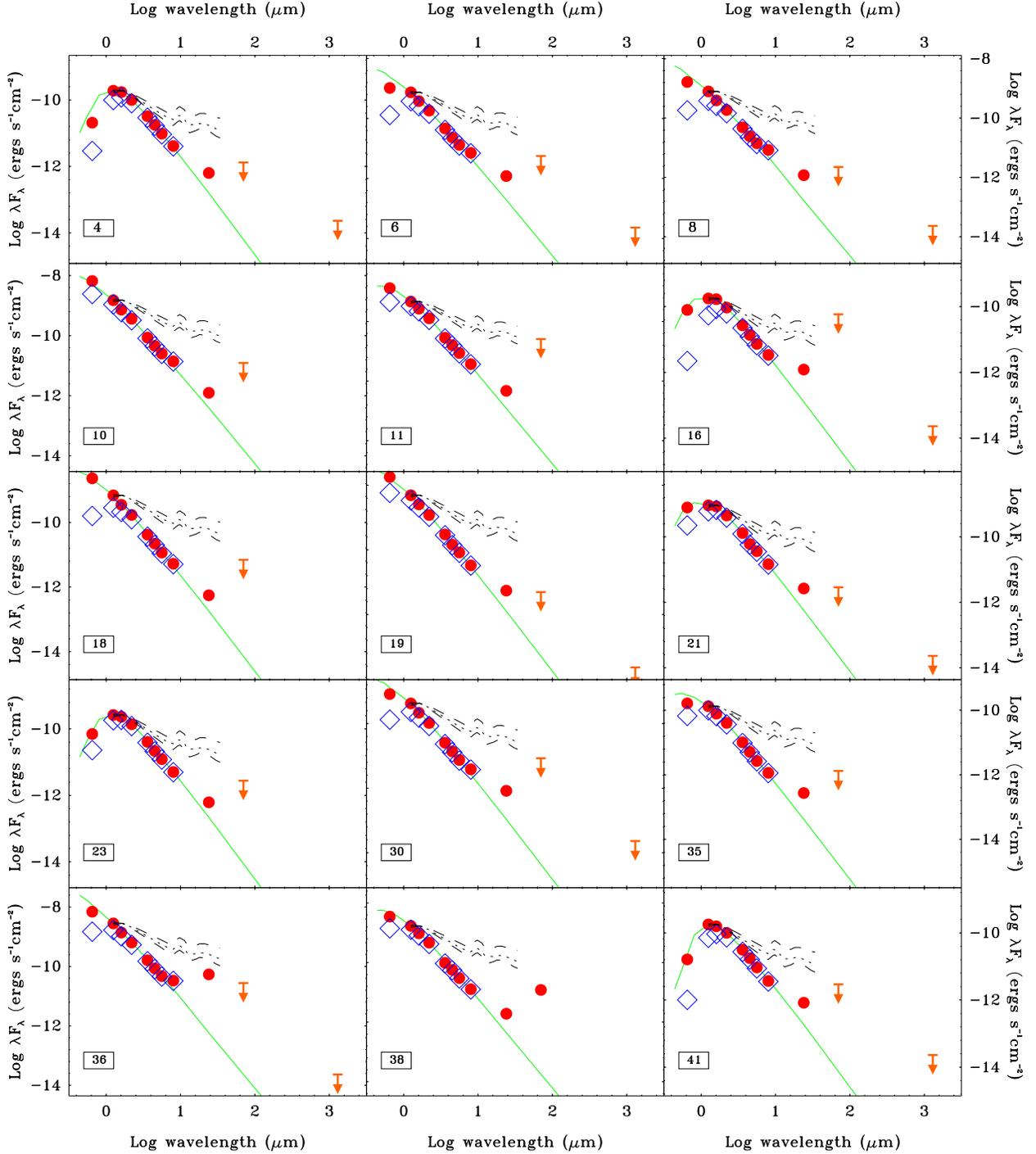}
\caption{The SEDs of  the 15 debris disk candidates. All of them  are non-accreting, have low disk masses ($\lesssim$ 5 M$_{JUP}$),
and/or fractional disk luminosities L$_{disk}$/L$_{star} <  10^{-3}$. 
As discussed in Section~\ref{pms_id}, objects \# 30, 35,  36, and 38 could be  debris disks around early-type background MS stars. 
}
\label{f:sed_debris}
\end{figure}

As discussed in Section~\ref{sed_mor}, 
typical primordial disks around CTTSs have L$_{D}$/L$_{*}$ $\sim$ 0.1 as they have 
optically thick disks that  reprocess $\sim$10$\%$ of the stellar radiation, while  
debris disks show  L$_{D}$/L$_{*}$  values $\lesssim$ 10$^{-3}$  because 
their optically thin disks only intercept $\sim$10$^{-5}$--10$^{-3}$ of the 
star's light (Bryden et al. 2006). 
Disks around young WTTSs have  L$_{D}$/L$_{*}$  values that fill the gap between those two regimes 
suggesting they are an evolutionary link between the two stages 
(Padgett et al. 2006; Cieza et al. 2007; Wahhaj et al. 2010). 
Following Papers I and II, we classify as photoevaporating disks objects with 
L$_{D}$/L$_{*}$~$\gtrsim$~10$^{-3}$ and as debris disks objects with
L$_{D}$/L$_{*}$~$\lesssim$~10$^{-3}$.
The SEDs of our 7 photevaporating disk candidates are shown in Figure~\ref{f:sed_photo}, while 
the SEDs of our 15 debris disk candidates are shown in Figure~\ref{f:sed_debris}.
The central objects of  4 of the debris disk candidates appear to be 
significantly  underluminous in the H$-$R diagram (targets \# 30, 35, 36, and 38)
and could be background MS stars (see Section~\ref{pms_id}).  
CO observations with the Atacama Large Millimeter/Submillimeter Array (ALMA) will  
be able to measure the gas content of all non-accreting disks and conclusively establish
their primordial or debris disk status.

\subsubsection{Circumbinary disks} 

Early multiplicity surveys of  PMS stars have shown that most  stars in the solar neighborhood form in multiple
systems (Leinert et al. 1993; Ghez et al. 1993; Simon et al. 1995). Since most of these binary systems have 
orbits with semimajor axes of the order of the typical sizes of circumstellar disks, the dynamical interaction of 
the stellar components with each other's disks has dramatic effects on disk devolution.
The outer disks around the individual stars in a binary system are expected to be tidally truncated at a fraction ($\sim$0.5) of the binary
separation.  Similarly,  the circumbinary disk, if present, should have an inner radius  $\sim$2$\times$ the semi-major axis 
of the system  (Artymowicz \& Lubow, 1994). 
While it is not the only possible outcome, tidal truncation in binary systems is known to produce inner opacity holes resulting in transition 
disk SEDs. Such is the case of the famous CoKu~Tau/4 system (Ireland $\&$ Kraus, 2008). 

In Section~\ref{multi_sec}, we found that 13 of our targets are in fact multiple systems.
We now discuss the likelihood  of each system to retain a circumbinary disk responsible 
for the observed SEDs. 
Nine of  the binary systems in our sample (objects \# 4, 5, 14, 15, 18, 22, 28, 31, and 32) 
have \emph{projected} separations greater than 100 AU (see Table 3). These projected 
separations represent the minimum value of 
the current physical separation.  Any circumbinary disk  around such wide systems 
should have an inner hole $\gtrsim$ 200 AU and would most likely remain 
undetectable  in our SEDs. 
Targets \#1,  24,  and 33 have companions at 62,  90, and 10 AU, respectively. 
However, they are all M-type stars with significant excesses at 8 $\mu$m micron,
implying the presence of dust at  separations of the order of a few AU or less. 
Also, their SEDs show no evidence for dynamically induced inner holes.  
We thus conclude that their transition disk status is not a result of their stellar 
companions. 
Object \# 26  (FW Tau), on the other hand, has a projected separation of only 7 AU and a
weak 24~$\mu$m excess and is therefore fully consistent with a circumbinary disk. The FW Tau system 
shares some of  the properties of Coku~Tau/4, such as its very low disk mass ($\lesssim$ 0.5 M$_{JUP}$) 
and  the lack of detectable accretion.  The lower luminosity of FW Tau (it is a M5 star) might explain
its lower mid-IR excess with respect to CoKu Tau/4 even if their inner holes are  of similar size.  We classify FW Tau as both a circumbinary and 
a photoevaporating disk candidate as tidal truncation and photoevaporation are not mutually exclusive processes.
An important  caveat of the circumbinary disk classification is the lack of constraints for most of the sample on stellar 
companions with separations $\lesssim$10 AU,  a range where $\sim$30$\%$ of all stellar companions are 
expected to be found (Duquennoy $\&$ Mayor, 1991;  Kraus et  al. 2011).  Future radial velocity and aperture masking observations are likely 
to increase the number of objects in the circumbinary disk category.

 \subsection{Implication for disk evolution} 
 
Combining our sample of  31 transition disks  with those from Papers I and II results in a
sample of 74 homogenously  selected and characterized transition disk  objects.  In this section, 
we discuss  the properties of the combined sample and the important clues transition disks can provide 
on different aspects of disk evolution.

\subsubsection{The incidence of planet-forming disk candidates}

We find that only $\sim$18$\%$ (13/74) of the transition disks in our combined 
sample have properties that are best explained by the dynamical interaction of 
recently formed giant planets. Calculating the overall incidence of planet-forming 
disk candidates among protoplanetary disks in nearby molecular clouds is non-trivial 
because  the fraction of transition disks and AGB contamination vary  greatly from 
cloud to cloud. We calculate this important quantity as follows: we exclude from our 
calculation the Taurus molecular cloud because its catalog does not provide a 
YSOc classification, which results in  planet-forming disk fraction of  21.0$\%$ (12/57) 
among the remaining well-characterized transition disks. From Table 3 in Paper II, 
we find that these 57 objects have been drawn from a sample of 1059 YSOc in 8 different 
molecular clouds (Lup I, III, IV, V and VI,  Ophiuhcus, Perseus, and Auriga), of which 
24.8$\%$ (263/1059) satisfy our  main sample selection criteria  ([3.6]-[4.5]  $<$ 0.25 
and [3.6]-[24] $>$ 1.5). We thus estimate the overall  fraction of planet-forming disk 
candidates among YSOc to be 0.210$\times$0.248 = 5.2$\%$.  
The YSOc catalogs  are known to be  contaminated by AGB stars,  and this contamination 
is higher among  objects with transition disk SEDs (AGB  typically have photosphere 
near-IR colors and small 24 $\mu$m excesses).   From the estimates in Paper II's Table 3, 
we find that  the overall YSOc contamination by AGBs in the 8 clouds we consider 
is 10.3$\%$, while the contamination increases to 15.6$\%$ in the region of the color-color 
diagram occupied by  transition disks. Correcting the statistics from AGB contamination 
decreases the incidence of transition disks in the YSOc sample to 23.3$\%$ (222/950).
Similarly,   the overall  fraction of planet-forming disk candidates among YSOc 
decreases slightly to  0.210$\times$0.233 = 4.9$\%$.

%
%

For comparison, the incidence of giant planets  within 20 AU of mature solar-type stars is estimated to be $\sim$20$\%$
(Cumming et al. 2008). This discrepancy strongly suggests that not all giant planets embedded in a primordial
disk result in clear observational signatures. 
The presence of  ``hidden planets" is supported by the results from recent hydrodynamic simulations showing
that multiple giant planets are needed in order to  have a detectable effect in the emerging SED
(Zhu et al. 2011; Dodson-Robinson \& Salyk 2011).
Furthermore, high-resolution (sub)millimeter images have revealed inner holes in systems  
that have perfectly ``normal'' SEDs (Andrews et al. 2011). 
This can easily be understood considering that (sub)millimeter observations can detect  modest reductions in the surface 
density of the disk,  while reducing the levels of the near and mid-IR excess emission requires an extreme 
depletion of  small dust particles. 
Also, the planet formation signature  capture by our selection criteria is expected to be present for only a small fraction
of the disk lifetime, after the planets have become massive enough to dynamically open a gap in the disk and the inner disk
has drained. 
In other words, many protoplanetary disks with ``normal" SEDs  may contain one or more protoplanets that are not 
yet massive enough  to open a wide hole in  the disk, and as mentioned in Section~2, we are likely to miss 
pre-transition disks, with wide gaps but significant near-IR excess. 
We thus conclude that  the giant planet disk candidates identified in our survey are likely to represent only a subsample of 
the entire population of disks actively forming planets in the molecular clouds we are considering.
This subsample may be dominated by systems with multiple and/or very massive planets.

\subsubsection{The incidence of  circumbinary disks}

Dynamical clearing by stellar companions has been one of the 
main mechanisms proposed to explain the inner holes of transition disks.
However, in Section~\ref{classification} we only found one circumbinary disk
candidate in our sample of 31 transition disks.  A low incidence of circumbinary disk
candidates was also found in Papers I and II.
The combined sample of 74 transition disks now allow us to derive more robust conclusions
on the role  multiplicity plays on transition disk systems. 
As in Paper I, we compare the distribution of binary separations of our sample
of transition disk to that of non-transition disk and disk-less PMS stars.
We draw our sample of non-transition disks (objects with ``normal" levels of near-IR excess) 
and disk-less PMS stars from the compilation of multiplicity and \emph{Spitzer} data (for disk identification)
presented by Cieza et al. (2009)  for over 300 PMS stars in the Taurus,  Chameleon I,  Ophiuchus, and Corona Australis regions.

\begin{figure}[t]
\includegraphics[width=5.5in]{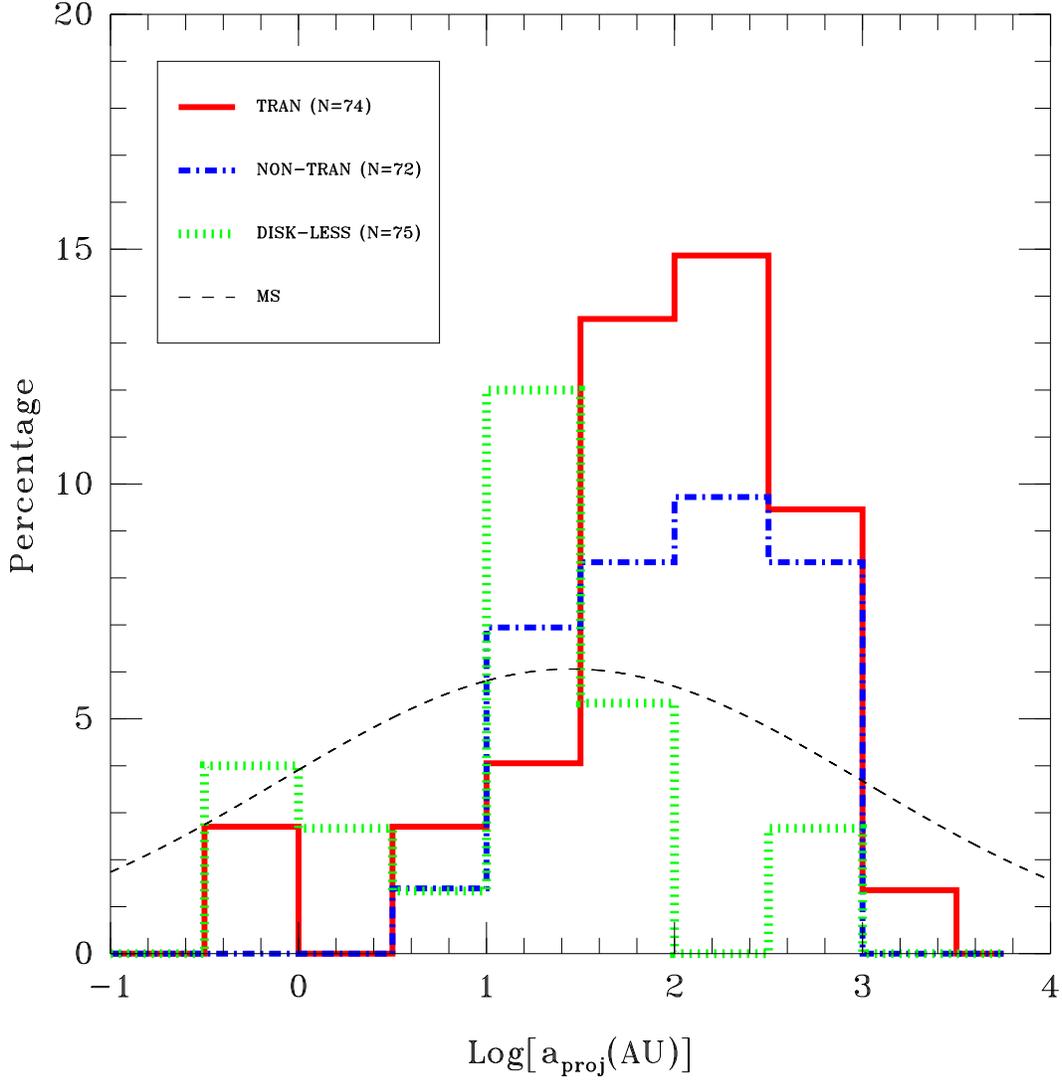}
\caption{Distribution of \emph{projected} companion separations for our combined transition disk sample (red solid line),  non-transition
disks from Cieza et al. 2009  (blue dash-doted line), and disk-less stars (green dotted line) also from
Cieza et al.  (2009). Spectroscopic binaries  have been assigned a projected separation of 0.5 AU. The total 
number of objects (single + multiple systems)  in each sample are shown in parenthesis.  The distribution of 
binary separations for main sequence (MS) solar-type stars (Duquennoy $\&$ Mayor, 1991) is shown for comparison.
Binary systems with separations in the 10 to 30 AU range result in the rapid erosion of the individual circumstellar disks.
Few circumbinary disks survive in such systems.
}
\label{f:multi_fig}
\end{figure}

As seen in Figure~\ref{f:multi_fig},  disk-less PMS stars tend to have companions
at smaller separations than stars with regular, non-transition disks.
According to a two-sided Kolmogorov--Smirnov (KS) test, there is less than a  10$^{-4}$ probability that the distributions 
of binary separations of non-transition disks and disk-less stars have been drawn from the same parent population.
This result can be understood in terms of the effect tidal truncation has on the  lifetimes of the circumstellar
disks of the individual components of the binary system (Cieza et al. 2009; Kraus et al. 2011).
A binary system at the peak of the separation distribution ($\sim$30 AU), is expected to initially 
have individual disks that are $\sim$10-15 AU in radius. Given that the viscous timescale is roughly 
proportional to the size of the disk, such small disks are likely to have accretion lifetimes smaller than the 
age of the sample and hence they now appear as young disk-less stars.
In very tight binary systems, the outer disk can survive in the form of a circumbinary disk, 
with a tidally truncated inner hole with a radius $\sim$2$\times$  the orbital separation 
(Artymowicz $\&$ Lubow, 1994).
If such close systems were a significant component of the transition disk population, 
one would expect to find a higher incidence of close binaries in transition disks that in 
non-transition disks.
While our multiplicity census is clearly very incomplete at small separations, 
our AO observations should be sensitive to an over abundance of companions
in the 10 to 30 AU range that could in principle be responsible for many of the inner
holes of our transition disks. However, we find that only $\sim$4$\%$ (3/74)
of our transition disks have companions in this separation range. 
Near-IR interferometry (Pott et al. 2010)  and aperture masking observations
(Kraus et al. 2009)
 also suggest that the incidence of tight \emph{stellar}
companions is rather low in transition objects.
We thus conclude that  tight stellar companions  tend to 
destroy each other's  disk rather quickly. Few close binary systems retain circumbinary disks 
with transition disk SEDs (e.g., CoKu Tau/4 and FW Tau). 

\subsubsection{Constrains on photoevaporation models}\label{photo_constrain}

While it is increasingly clear  that photoevaporation by the central star
plays a fundamental role on  the evolution and dissipation of protoplanetary disks,
the heating  mechanisms and magnitudes of the photoevaporation rates are still
a matter of intense debate. In particular, the relative importance of  Far-UV (FUV), Extreme-UV (EUV),
and X-ray photoevaporation is not well understood.  All photoevaporation models
predict the formation of an inner hole and subsequent
inside-out dissipation once mass transport across the disk falls below the
photoevaporation rate; however,  they strongly disagree on the mass a disk should have at the time
 the inner hole is formed. EUV-driven photoevaporation models (Alexander et al. 2006) predict
low evaporation rates, of the order of 10$^{-10}$ M$_{\odot}$/yr. As a result, the inner hole
forms late on the evolution the disk, when most of the mass has been depleted (disk mass $\lesssim$ 1 M$_{JUP}$)
and the accretion rate onto the star has dropped below the 10$^{-10}$ M$_{\odot}$/yr level.
More recent EUV+X-ray and FUV+X-ray  photoevaporation models
(Gorti et al. 2009; Owen et al. 2011)
predict  much higher evaporation rates (10$^{-9}$--10$^{-8}$ M$_{\odot}$/yr) and an earlier formation for
the inner hole, when disk masses are $\gtrsim$~5 M$_{JUP}$.
These later models allow the presence of  accretion in disks with inner holes that have formed through 
photoevaporation because the accretion rates from the draining of the inner disk 
become detectable for significant  period of time. 
However, once the inner disk is drained, accretion must stop as  material from the \emph{outer} disk  can not 
overcome the photoevaporation front. 
X-ray photevaporation models  thus predict a population of non-accreing objects (i.e., WTTSs)  with 
relatively massive outer disks ($\gtrsim$~5~M$_{JUP}$).

\begin{figure}[t]
\includegraphics[width=7.0in]{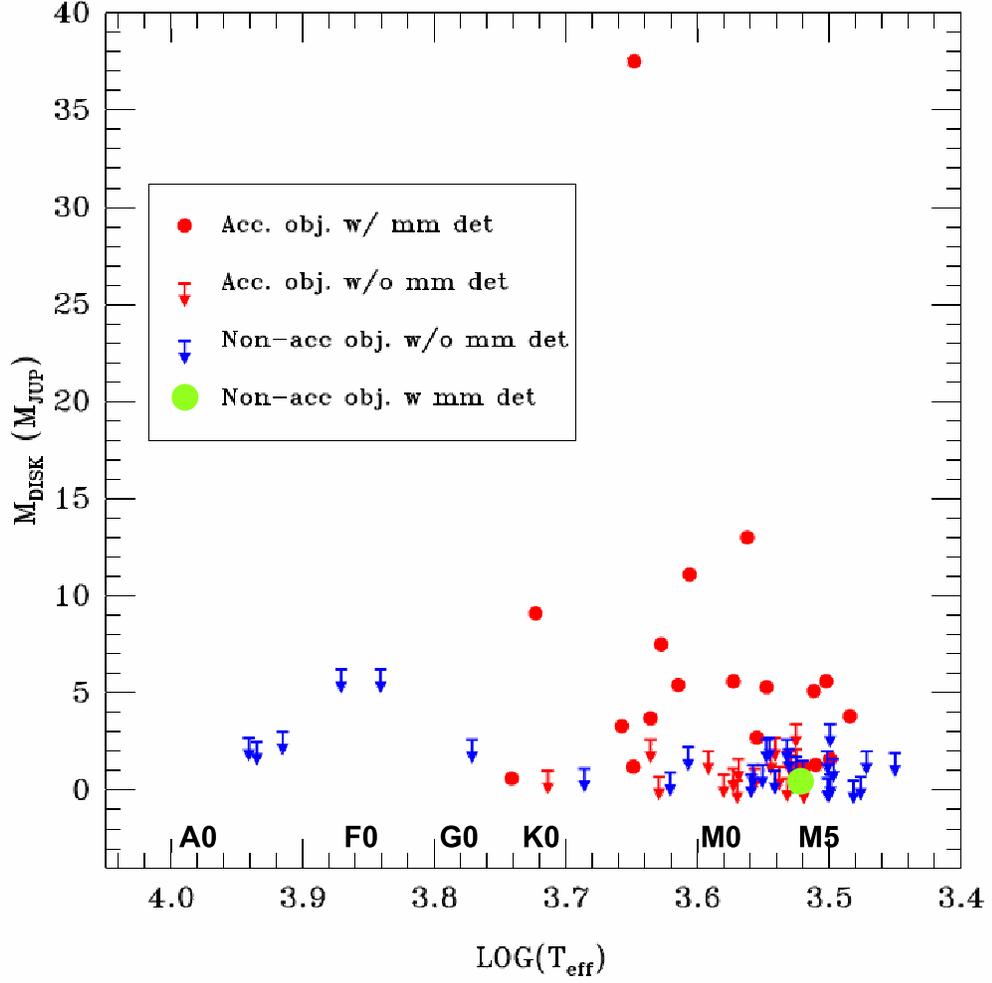}
\caption{Disk mass as function of stellar spectral type for our combined sample of transition disks.
To avoid overlapping data points, small random offsets have been applied in the X-axis.
Only one non-accreting object has been detected at millimeter wavelengths, FW Tau, with
an estimated disk mass of  0.4 M$_{JUP}$. The lack of non-accreting objects with relative massive
outer disks ($\gtrsim$ 2--5 M$_{JUP}$) favors photoevaporation models with low evaporation rates ($\sim$10$^{-10}$M$_{\odot}$yr$^{-1}$)
across a wide range of stellar masses. 
}
\label{f:mass_vs_spt}
\end{figure}

In Paper II, we demonstrated the lack of massive  outer disk  around low-mass WTTSs,  which contradicts the predictions 
of photoevaporating models  with high accretion rates. 
This result  stills holds for the new low-mass stars studied herein and seems to extend to higher mass objects. 
Figure~\ref{f:mass_vs_spt} shows the disk masses derived for our combined sample of transition disks. 
There is only one (sub)millimeter detection at 0.4 M$_{JUP}$ for non-accreting objects (corresponding to FW Tau), 
the rest are  upper limits in the $\sim$0.5 to $\sim$5 M$_{JUP}$ range.  Since FW Tau is a circumbinary disk,  there is not
guarantee that its inner hole is due to photoevaporation.  
The sample now includes several non-accreting KGFAB stars observed at millimeter
wavelengths, but increasing the sample size for this type of objects is clearly desirable. 
An important caveat to Figure~\ref{f:mass_vs_spt}  is the inherent uncertainty of deriving disk masses
from (sub)millimeter photometry, which requires strong assumptions on 
dust opacities and gas to dust mass ratios (Williams $\&$ Cieza, 2011).  If those assumptions are incorrect, 
they could lead to systematic errors in the determination of disk masses.
However, the observed accretion rates of PMS stars  provide independent observational constraints that also
favor photoevaporations rates significantly lower than 10$^{-8}$ M$_{\odot}$/yr. If photoevaporation rates are $\sim$10$^{-8}$ M$_{\odot}$/yr,
then it becomes difficult to explain all the CTTSs without inner holes that have accretion rates less than $\sim$10$^{-8}$ M$_{\odot}$/yr
(Hartmann et al. 1998).
Overall, the observational evidence suggests that photoevaporation rates must be small ($\sim$10$^{-10}$ M$_{\odot}$yr$^{-1}$)
 and that the phoptoevaporation front  can only overcome accretion when disk masses have fallen below 
 $\lesssim$~1~M$_{JUP}$.

\subsubsection{Disk types as a function of stellar spectral type and age}\label{mass-dep} 

\begin{figure}[t]
\includegraphics[width=7.0in]{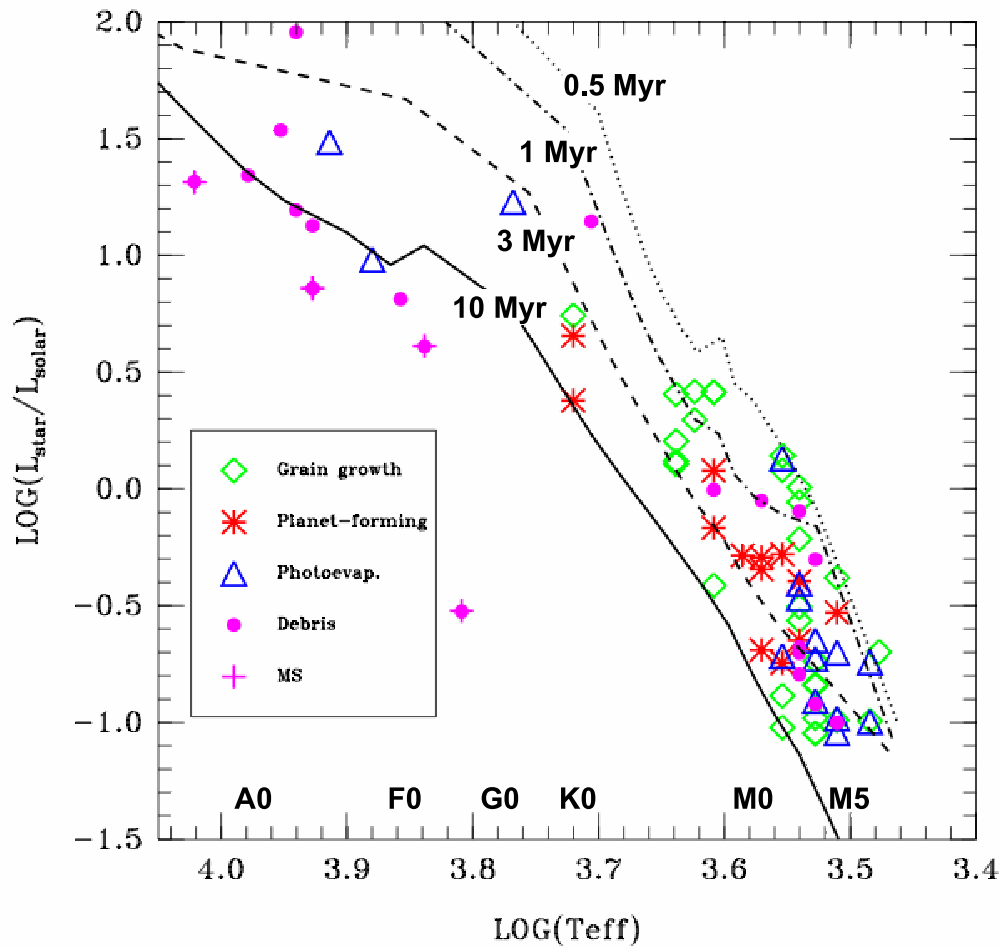}
\caption{
The  H-R diagram as in Figure~\ref{f:HRD}, but showing the location of each type of disk for the 
combined sample of disks from Papers I, II, and III.
All  stars hotter than $\sim$10$^{3.76}$ (5754 K, corresponding to 
a G5 star) have non-accreting disks, either photoevaporating disks or debris disk,
consistent with the idea that primordial disks dissipate faster around  more massive objects.
There is a lack of (giant) planet-forming disk candidates  among the youngest stars in the sample.
This favors core accretion as the main planet formation mechanism and a 2 to 3 Myr formation timescale.
}
\label{f:HRD_types}
\end{figure}

While the samples in  Papers I and II  were strongly dominated by M-type stars, our current combined sample 
does contain a significant number of higher mass KGFAB-type stars. 
These allow us to investigate whether the incidence of the 
disk clearing mechanisms depend on stellar mass or luminosity.
Figure~\ref{f:HRD_types} shows the same H-R diagram as in Figure~\ref{f:HRD}, 
but now indicating the location of each type of disk. 
The most striking feature of the diagram is that all 9 PMS stars hotter than $\sim$10$^{3.76}$ (5754 K, corresponding to 
a G5 star) have non-accreting disks, either photoevaporating disks or debris disks. 
This is  in agreement with the results from \emph{Spitzer} studies of young stellar clusters and associations 
showing that primordial disks dissipate faster around higher mass stars (Carpenter et al.  2006; Dahm $\&$ Hillenbrand,  2007).
The 4 MS stars with debris disks are also hot BAF-type stars.
This can be understood considering that debris disks are  easier to detect at mid-IR wavelengths around 
more luminous objects (e.g., Rieke et al. 2005; Hernandez et al.  2006; Cieza et al. 2008a).
Table~4 shows the occurrence of  different types of transition disks
for M-type (48 objects), K-type (17 objects) and GFAB-type stars (9 objects) in
our combined sample.  
Since M-type stars are fainter and cooler than  higher mass stars, they may present
weaker near- and/or mid-IR excesses even without grain growth and dust settling effects  
(Ercolano, Clarke $\&$ Robitaille, 2009). 
However, the fact that we see a  \emph{lower} fraction of grain-growth 
dominated disk candidates around M-type stars than around 
K-type stars,  suggests that we are not significantly overestimating the fraction  of
M-type stars in this category. 
Detailed modeling of individual objects are still needed to
confirm the nature of some of these grain-growth dominated disks, especially the ones
around very late M-type stars with modest decrements of IR emission. 
However,  it is clear that grain-growth and dust settling 
play a fundamental role on the evolution of cicumstellar disks.
The grain-growth dominated disk category accounts for $\gtrsim$40$\%$
of all disks around both the M-type stars and K-type stars in our sample.

Currie $\&$ Sicilia-Aguilar (2011) recently showed that the percentage 
of disks in the transition phase increases significantly between $\sim$1 to $\sim$8 Myr. 
The H-R diagram in Figure~\ref{f:HRD_types}, showing our combined sample, 
now allows us to investigate how the incidence of different transition disk \emph{types} 
evolve with age. 
While the stellar ages of individual targets are highly uncertain, the age distribution of each disk
category should carry more meaningful information. 
Table~5 shows the occurrence of  different types of transition disks
for objects falling above the 1 Myr isochrone (14 targets), between the 1 and 3 Myr isochrones (30 objects), 
and  below the 3 Myr isochrone (30 objects, \emph{not} including the 4 MS star candidates) in our combined sample. 
Even with our relatively small sample,  we find that the incidence of photoevaporating disks and debris disks 
increases with age, while the fraction of  grain-growth dominated disks decreases with time.
These trends are in agreement with the overall evolution of typical protoplanetary disks
(Williams $\&$ Cieza, 2011). 
Interestingly, the occurrence of plant-forming disks candidates  peaks in the 1-3 Myr old age bin
and is $\sim$0$\%$ for objects above the 1 Myr age isochrone. 
The lack of planet-forming disks with ages $\lesssim$ 1 Myr is suggestive, especially considering that
the intrinsic  age distribution of these objects  is most likely to be narrower 
than that seen in the H-R diagram, where some of the scatter can be attributed to the observational
uncertainties  in T$_{eff}$ and luminosity. 
If taken at face at value, these results would imply that \emph{1)}  the inferred giant planets have
formed through core-accretion (Lissauer, 1993) as gravitational instability models (Boss, 2000)
favor a younger age distribution, and \emph{2)} core accretion takes 2 to 3 Myr to form giant planets 
massive enough to open a gap in the disk. 
While important caveats remain (we are dealing with planet-forming disk \emph{candidates}, 
the sample is relatively small, and stellar ages are highly uncertain and model dependent), 
these results are promising. 
In the near future, detailed age analyses of larger samples of transition disks with 
increasingly clear planet-formation signatures are likely to provide strong
astrophysical constraints on planet-formation timescales.

\section{Summary and Conclusions}

As part of an ongoing program aiming to characterize a large number
of  \emph{Spitzer}-selected transition disks,  we have obtained millimeter 
wavelength photometry, high-resolution optical spectroscopy and adaptive 
optics near-infrared imaging for a sample of  31  transition objects located 
in the Perseus, Taurus, and Auriga molecular clouds.
We use these ground-based data to estimate  disk masses,  
multiplicity, and accretion rates in order to  investigate the mechanisms 
potentially responsible for  their inner holes.
We combined disk masses, accretion rates, and multiplicity data with other information, such as 
SED morphology and fractional disk luminosity to classify the disks as \emph{strong candidates} 
for the following categories: grain-growth dominates disks (7 objects),  giant planet-forming disks (6 objects), photoevaporating 
disks (7 objects),  debris disks (11 obecjts), and cicumbinary disks (1 object, which was also classified as photoevaporating disk). 
Each category represents an educated guess, giving all the available data, on the evolutionary status of
the disk or the physical process mainly responsible for the reduced levels of near-IR and/or mid-IR excesses
characteristic of the objects in our sample.  
The boundaries between the categories are of course not perfectly defined. For instance, all primordial
disks are expected to simultaneously undergo some degree of grain growth, dust settling, and
photoevaporation.  Similarly, our criterion to distinguish primordial photoevaporating disks from debris disks
is based on L$_{D}$/L$_{*}$  instead of on the gas content of their outer disks, which still remains
highly unconstrained.  
The gas dissipation of the outer disk through photoevaporation is believed to mark the rapid transition from the primordial to the 
debris disk stage (Williams $\&$ Cieza, 2011). 
Conclusively establishing the nature of each target will thus require detail modeling and followup observations.  
Even with the above caveats,   the properties of transition disks can provide important clues on different  aspects 
of disk evolution.
Combining our sample of  31 transition disks with those from Papers I and II results in a
sample of 74 transition disk  objects that have been selected, characterized,  and classified in
an homogenous way. 
The main conclusions derived from the analysis of this combined high-quality sample can
be summarized as follows: 

\noindent 1)   Circumstellar disks with reduced levels of near-IR and/or mid-IR excesses
in nearby molecular clouds represent a very heterogenous group of objects with a wide
range of SED morphologies, disk masses,  accretion rates, and fractional disk luminosities.
This diversity points toward distinct evolutionary stages and physical 
processes driving the evolution of each disk. 

\noindent 2) The incidence of objects with signatures of dynamical clearing by recently formed
giant planets is significantly lower than the occurrence of giant planets  within 
20 AU of mature solar-type stars ($\sim$5$\%$ vs $\sim$20$\%$).
The giant planet disk candidates identified in our survey are likely to represent  special cases,  
where  multiple massive planets may be present. 

\noindent 3) The  incidence of circumbinary disk candidates in our sample of transition 
objects is low ($\lesssim$ 10$\%$) implying that  tight \emph{stellar} companions  tend to 
erode each other's  disk rather quickly. 

\noindent 4) There is a lack of massive disks around non-accreting stars
of a wide range of spectral types,  which contradicts the predictions of  recent photoevaporation models
that find very high evaporation rates  ($\sim$10$^{-8}$ M$_{\odot}$yr$^{-1}$).
Our results suggest  that photoevaporation rates must be small 
($\sim$10$^{-10}$ M$_{\odot}$yr$^{-1}$) and that the photoevaporation front  can 
only overcome accretion when disk masses have fallen below  $\lesssim$ 1 M$_{JUP}$. 

\noindent 5) Debris disks and photoevaporating disk candidates are more common around hotter
stars, consistent with the idea that primordial disks dissipate faster around  more massive objects.

\noindent 6) Grain growth-dominated disks account for $\gtrsim$40$\%$ of our sample of 
transition disks around K and M-type stars, confirming that grain-growth and dust settling play a 
major role on the evolution of primordial circumstellar disks. 

\noindent 7) We find a trend in the sense that the incidence of photoevaporating disks and debris
disks increases with age, while the fraction of grain-growth dominated disks decreases with
time, which is consistent with disk evolution models.

\noindent 8)  A preliminary analysis of the age distribution of disks  with signatures of dynamical clearing by recently formed
giant planets  reveals a lack of such objects among the youngest stars in the sample. This favors core accretion 
as the main planet formation mechanism and a 2 to 3 Myr formation timescale.

Transition objects are invaluable disk evolution and planet formation
laboratories. Detailed modeling and followup observations of different types of transition disks 
are highly desirable to further our  understanding of key processes such as 
grain growth and dust settling,  photoevaporation, dynamical clearing, 
and planet formation itself.

\acknowledgments
{\it{Acknowledgments:}} 

We thank the anonymous referee whose comments and suggestions
helped us to significantly improve the paper.
Support for this work was provided by NASA through
the \emph{Sagan} Fellowship Program under an award from Caltech.
M.R.S thanks for support from FONDECYT (1061199) and Basal CATA PFB 06/09.
G.A.R. was supported by ALMA FUND Grant 31070021. M.D.M. was supported by ALMA-Conicyt
FUND Grant 31060010.
J.P.W. acknowledges support from the National Science Fundation Grant AST08-08144.
ARM acknowledges financial support from Fondecyt in the form of grant  number  3110049.
This work makes use of  data obtained with the \emph{Spitzer} Space Telescope,
which is operated by JPL/Caltech,  under a contract with NASA. 

\emph{Facilities}:  {\it{Spitzer}} (IRAC, MIPS), SMA,  JCMT (SCUBA-2), Gemini  (NIRI), and
CFHT (Espadons).

\begin{deluxetable}{rrcrrrrrrrrrrr}
\rotate
\tablewidth{0pt}
\tabletypesize{\tiny}
\tablecaption{Transition Disk Sample}
\label{sample}
\tablehead{\colhead{\#}&\colhead{2MASS ID}&\colhead{Alter. Name}&\colhead{R}&\colhead{J\tablenotemark{a}}&\colhead{H}&\colhead{K$_{S}$}&\colhead{F$_{3.6}$\tablenotemark{a}}&\colhead{F$_{4.5}$}&\colhead{F$_{5.8}$}&\colhead{F$_{8.0}$}&\colhead{F$_{24}$}&\colhead{F$_{70}$\tablenotemark{b}}&\colhead{Region}    \\
\colhead{}&\colhead{}&\colhead{}&\colhead{(mag)}&\colhead{(mJy)}&\colhead{(mJy)}&\colhead{(mJy)}&\colhead{(mJy)}&\colhead{(mJy)}&\colhead{(mJy)}&\colhead{(mJy)}&\colhead{(mJy)}& \colhead{(mJy)}& \colhead{}}
\startdata
  1 & 03292681+3126475 &  MBO 15                     &        12.37 & 7.41e+01 & 1.03e+02 & 8.96e+01 & 4.72e+01 & 3.20e+01 & 2.33e+01 & 1.70e+01 & 1.89e+01 & $<$     8.06e+01 & PER  \\
  2 & 03292925+3118347 &  MBO 22                     &        16.56 & 1.47e+01 & 2.89e+01 & 2.75e+01 & 1.36e+01 & 1.02e+01 & 7.62e+00 & 6.94e+00 & 9.52e+01 &             2.91e+02 & PER  \\
  3 & 03302409+3114043 &  [EDJ2009] 261         &        17.30 & 6.83e+00 & 7.73e+00 & 6.93e+00 & 4.99e+00 & 3.96e+00 & 3.80e+00 & 5.04e+00 & 5.43e+00 & $<$     1.05e+01 & PER  \\
  4 & 03335108+3112278 &  [EDJ2009] 300         &        16.72 & 4.15e+01 & 6.37e+01 & 5.87e+01 & 3.50e+01 & 2.50e+01 & 1.72e+01 & 1.07e+01 & 5.06e+00 & $<$     1.53e+01 & PER  \\
  5 & 03344987+3115498 &  [EDJ2009] 303         &        13.79 & 8.34e+01 & 1.18e+02 & 1.04e+02 & 6.71e+01 & 4.57e+01 & 3.44e+01 & 3.43e+01 & 4.86e+01 &             4.04e+01 & PER  \\
  6 & 03411412+3159462 &  [EDJ2009] 307         &        12.59 & 1.57e+02 & 1.37e+02 & 1.06e+02 & 5.00e+01 & 3.24e+01 & 2.34e+01 & 1.85e+01 & 9.64e+00 & $<$     6.53e+01 & PER  \\
  7 & 03413918+3136106 &  BD+31 634               &        10.00 & 2.37e+02 & 1.68e+02 & 1.16e+02 & 5.05e+01 & 3.61e+01 & 2.99e+01 & 5.26e+01 & 1.51e+02 & $<$     5.75e+01 & PER  \\
  8 & 03422333+3157426 &  [EDJ2009] 323         &        12.23 & 1.60e+02 & 1.39e+02 & 1.07e+02 & 5.22e+01 & 3.39e+01 & 2.50e+01 & 2.19e+01 & 9.45e+00 & $<$     2.61e+01 & PER  \\
  9 & 03434461+3208177 &  Cl* IC 348 LRL 67   &        14.65 & 2.41e+01 & 3.61e+01 & 3.21e+01 & 1.90e+01 & 1.49e+01 & 1.14e+01 & 1.09e+01 & 1.04e+02 &             1.73e+02 & PER  \\
10 & 03440915+3207093 & Cl* IC 348 LRL 8      &           9.41 & 4.55e+02 & 3.24e+02 & 2.39e+02 & 9.79e+01 & 6.64e+01 & 4.63e+01 & 3.68e+01 & 1.00e+01 & $<$     1.44e+02 & PER  \\
11 & 03441912+3209313 & Cl* IC 348 LRL 30    &        11.24 & 1.37e+02 & 1.18e+02 & 8.32e+01 & 3.24e+01 & 2.33e+01 & 1.60e+01 & 9.81e+00 & 3.78e+00 & $<$     3.02e+02 & PER  \\
12 & 03442156+3215098 & Cl* IC 348 LRL 185 &      \nodata & 1.01e+01 & 1.31e+01 & 1.15e+01 & 5.66e+00 & 4.29e+00 & 3.04e+00 & 1.99e+00 & 4.84e+00 & $<$     1.71e+02 & PER  \\
13 & 03442257+3201536 & Cl* IC 348 LRL 72    &        15.93 & 2.26e+01 & 3.55e+01 & 3.23e+01 & 1.72e+01 & 1.11e+01 & 8.91e+00 & 7.55e+00 & 6.76e+01 & $<$     7.36e+02  & PER  \\
14 & 03443200+3211439 & Cl* IC 348 LRL 12B &        12.88 & 1.49e+02 & 1.84e+02 & 1.80e+02 & 1.38e+02 & 1.08e+02 & 2.61e+02 & 7.66e+02 & 3.68e+02 & $<$     8.65e+02 & PER  \\
15 & 03443694+3206453 & Cl* IC 348 LRL 6      &        11.69 & 3.31e+02 & 4.13e+02 & 3.52e+02 & 2.08e+02 & 1.48e+02 & 1.16e+02 & 9.26e+01 & 4.61e+01 & $<$     5.38e+02 & PER \\
16 & 03444351+3207427 &  Cl* IC 348 LRL 52   &        17.01 & 2.27e+01 & 4.53e+01 & 4.54e+01 & 2.68e+01 & 1.83e+01 & 1.26e+01 & 8.65e+00 & 9.77e+00 & $<$     6.80e+02 & PER  \\
17 & 03445614+3209152 &  Cl* IC 348 LRL 21   &        14.81 & 6.21e+01 & 1.03e+02 & 1.08e+02 & 8.69e+01 & 6.53e+01 & 5.53e+01 & 4.38e+01 & 2.14e+02 & $<$     2.30e+02 & PER  \\
18 & 03450142+3205017 &  Cl* IC 348 LRL 25   &        12.38 & 1.19e+02 & 1.13e+02 & 9.24e+01 & 4.36e+01 & 3.01e+01 & 2.04e+01 & 1.36e+01 & 4.44e+00 & $<$     8.10e+01 & PER \\
19 & 04104210+3805598 &  BD+37 887               &         8.47 & 1.41e+03 & 1.10e+03 & 7.76e+02 & 3.41e+02 & 2.10e+02 & 1.51e+02 & 8.71e+01 & 4.39e+01 & $<$      5.77e+01  & AUR  \\
20 & 04190110+2819420 &  [GBA2007] 527        &        16.50 & 9.81e+01 & 1.48e+02 & 1.49e+02 & 8.79e+01 & 6.12e+01 & 4.42e+01 & 3.19e+01 & 2.43e+02 &            4.10e+02  & TAU \\
21 & 04192625+2826142 &  V819 Tau                  &        12.01 & 2.52e+02 & 3.56e+02 & 2.85e+02 & 1.47e+02 & 8.64e+01 & 6.58e+01 & 3.82e+01 & 2.11e+01 & $<$     3.35e+01  & TAU  \\
22 & 04210934+2750368 &  \nodata                      &        15.55 & 5.14e+01 & 5.59e+01 & 4.79e+01 & 2.90e+01 & 2.19e+01 & 1.66e+01 & 1.32e+01 & 9.46e+00 & $<$     2.25e+01 & TAU \\
23 & 04242321+2650084 &  \nodata                      &         14.48 & 7.57e+01 & 1.00e+02 & 8.94e+01 & 4.61e+01 & 3.13e+01 & 2.24e+01 & 1.33e+01 & 4.95e+00 & $<$     3.23e+01 & TAU \\
24 & 04284263+2714039 & WDS J04287+2714AB  &   16.81 & 2.29e+01 & 3.83e+01 & 4.39e+01 & 3.50e+01 & 2.76e+01 & 2.39e+01 & 1.89e+01 & 2.27e+01 & $<$     2.29e+01 & TAU  \\
25 & 04292083+2742074 & IRAS 04262+2735    &        15.54 & 6.06e+02 & 1.06e+03 & 1.04e+03 & 6.04e+02 & 4.17e+02 & 3.69e+02 & 5.95e+02 & 4.28e+02 & $<$     2.63e+01 & TAU \\
26 & 04292971+2616532 & FW Tau                        &        14.65 & 1.16e+02 & 1.38e+02 & 1.17e+02 & 6.48e+01 & 4.48e+01 & 3.24e+01 & 1.80e+01 & 6.79e+00 & $<$     2.35e+01 & TAU \\
27 & 04295531+2258579 &  IRAS 04269+2252   &        13.13 & 4.41e+03 & 8.11e+03 & 8.64e+03 & 2.70e+03 & 1.93e+03 & 1.82e+03 & 1.45e+03 & 5.17e+02 & $<$     1.94e+01& TAU  \\
28 & 04300113+3517247 &  HBC 390                     &       15.51 & 7.47e+01 & 1.37e+02 & 1.30e+02 & 7.57e+01 & 4.80e+01 & 3.08e+01 & 2.29e+01 & 2.65e+01 & $<$     1.43e+03 & AUR\\
29 & 04300424+3522238 &  \nodata                        &       16.53 & 8.21e+00 & 1.30e+01 & 1.13e+01 & 6.47e+00 & 4.25e+00 & 3.23e+00 & 2.69e+00 & 2.48e+01 &  $<$       3.54e+02 & AUR \\
30 & 04301644+3525217 &  [HAD2004] LDN 1482 G &12.58 & 9.60e+01 & 8.15e+01 & 6.20e+01 & 2.78e+01 & 1.93e+01 & 1.33e+01 & 1.00e+01 & 6.63e+00 & $<$     9.85e+01 & AUR\\
31 & 04303235+3536133 & \nodata                         &       15.94   & 3.13e+01 & 6.07e+01 & 6.08e+01 & 3.37e+01 & 2.42e+01 & 1.72e+01 & 1.55e+01 & 2.92e+02 & $<$     9.60e+01 & AUR\\
32 & 04304004+3542101 & \nodata                        &        15.76   & 1.83e+01 & 2.73e+01 & 2.49e+01 & 1.44e+01 & 1.02e+01 & 7.32e+00 & 5.57e+00 & 7.36e+00 & $<$     4.46e+01  & AUR \\
33 & 04305137+2442222 &  ZZ Tau                        &        13.31   & 2.54e+02 & 3.41e+02 & 2.80e+02 & 1.65e+02 & 1.24e+02 & 1.00e+02 & 1.09e+02 & 1.07e+02 & $<$     9.50e+01 & TAU \\
34 & 04312113+2658422 &  IRAS 04282+2652   &        14.76 & 3.14e+03 & 6.25e+03 & 7.24e+03 & 2.46e+03 & 1.92e+03 & 2.03e+03 & 1.29e+03 & 3.84e+02 & $<$     3.00e+01 & TAU \\
35 & 04314503+2859081 & \nodata                        &        13.32  & 4.14e+01 & 3.52e+01 & 2.68e+01 & 1.16e+01 & 7.45e+00 & 4.89e+00 & 3.03e+00 & 2.18e+00 & $<$     1.54e+01 & TAU \\
36 & 04330422+2921499 & BD+29 719                 &         9.96 & 7.02e+02 & 5.50e+02 & 3.93e+02 & 1.79e+02 & 1.22e+02 & 8.49e+01 & 8.72e+01 & 4.31e+02 & $<$     3.22e+02 & TAU \\
37 & 04343549+2644062 &   \nodata                      &        13.90 & 5.22e+02 & 1.11e+03 & 1.17e+03 & 6.57e+02 & 3.97e+02 & 3.02e+02 & 2.09e+02 & 8.01e+01 &$<$   1.28e+01 & TAU \\
38 & 04364912+2412588 & HD 283759                 &        10.01 & 5.27e+02 & 4.32e+02 & 3.09e+02 & 1.12e+02 & 8.25e+01 & 5.50e+01 & 3.36e+01 & 1.53e+01 &            2.80e+02 & TAU \\
39 & 04372486+2709195 &  HD 283751                &        11.25 & 1.94e+02 & 1.55e+02 & 1.27e+02 & 4.65e+01 & 3.41e+01 & 2.70e+01 & 2.07e+01 & 8.21e+00 & $<$     2.37e+01 & TAU \\
40 & 04385827+2631084 & Elia 3-14                      &        15.60 & 2.73e+02 & 8.58e+02 & 1.16e+03 & 7.48e+02 & 4.85e+02 & 3.82e+02 & 2.42e+02 & 9.58e+01 & $<$     3.24e+01& TAU  \\
41 & 04403979+2519061& WDS J04407+2519AB &     17.87 & 2.98e+01 & 4.94e+01 & 5.36e+01 & 3.27e+01 & 2.41e+01 & 1.63e+01 & 9.49e+00 & 6.64e+00 & $<$     3.41e+01 & TAU 
\enddata
\tablenotetext{a}{All the 2MASS, IRAC and 24 $\mu$m  detections are $\ge$7-$\sigma$ (i.e.,the photometric uncertainties are $\lesssim$15$\%$)} 
\tablenotetext{b}{$\ge$5-$\sigma$ detections or  5-$\sigma$ upper limits} 
\end{deluxetable}

\begin{deluxetable}{rrrlrrrrrrrr}
\tablewidth{0pt}
\tabletypesize{\scriptsize}
\tablecaption{Observed Properties}
\label{observed}
\tablehead{\colhead{\#}&\colhead{Ra (J2000)}&\colhead{Dec (J2000)}&\colhead{SpT.}&\colhead{H$\alpha$\tablenotemark{a}}&\colhead{$\lambda_{mm}$}&\colhead{Flux$_{mm}$\tablenotemark{b}}&\colhead{$\sigma$Flux$_{mm}$}&\colhead{Separ}&\colhead{pos. ang.}&\colhead{$\Delta$K} \\
\colhead{}&\colhead{(deg)}&\colhead{(deg)}&\colhead{}&\colhead{(km/s)}&\colhead{(mm)}&\colhead{(mJy)}&\colhead{(mJy)}&\colhead{(arcsec)}&\colhead{(deg)}&\colhead{(mag) }}
\startdata
  1  &  52.3617  &  31.4465  &  M2       &  140  & 1.30  &$<$    2.9 &   \nodata  & 0.25; 1.7 &   188; 63 &  0.83; 2.09         \\
  2  &  52.3720  &  31.3096  &  M0       &  280  &  1.30  &  6.3    &    1.1 & \nodata & \nodata &  $>$ 1.65; 3.85        \\
  3  &  52.6003  &  31.2345  &  M2      &  340  & 1.30   &$<$    3.0   &    \nodata & \nodata & \nodata &   $>$ 1.82; 3.70       \\
  4  &  53.4628  &  31.2077  &  M3      &  170  & 1.30   &$<$   2.9 &    \nodata & 0.96 &  101 &    2.08         \\
  5  &  53.7077  &  31.2640  &  M2       &  450  &  1.30   &6.0    &     1.2  &  0.85 &  231 &   2.62         \\
  6  &  55.3088  &  31.9962  & A3        &  -1     &  0.85  & $<$ 15.0    &    \nodata & \nodata & \nodata & $>$ 2.95; 5.23        \\
  7  &  55.4132  &  31.6030  & A5&  -1    &  1.30   &     $<$   3.0    &    \nodata & \nodata & \nodata & $>$  2.62; 4.86        \\
  8  &  55.5972  &  31.9619  & A0    &  -1    &  0.85 &   $<$ 15.0    &    \nodata & \nodata & \nodata & $>$  3.08; 4.73        \\
  9  &  55.9359  &  32.1383  & M1    &  280&  0.85  &31   &    6.0 & \nodata & \nodata & $>$   1.91; 4.01        \\
10  &  56.0382  &  32.1192  & A2 &  -1   &  \nodata  & \nodata&    \nodata & \nodata & \nodata & $>$   2.64; 4.33         \\
11  &  56.0797  &  32.1587  & F0 &  -1   &  \nodata  &  \nodata&    \nodata & \nodata & \nodata & $>$    2.77;  4.49        \\
12  &  56.0899  &  32.2527  & M2          &  130   &1.30   & $<$  3.0    &    \nodata & \nodata & \nodata & $>$  0.95;  3.14      \\
13  &  56.0941  &  32.0316  & M2   &  140   & 1.30   &$<$ 1.5    &    \nodata & \nodata & \nodata & $>$  1.23;  3.03         \\
14  &  56.1335  &  32.1955  & A5       &  -1   &  1.30  &$<$  2.8    &    \nodata &  1.31 & 274&    0.20        \\
15  &  56.1540  &  32.1126  & G2 &  -1   & 1.30   &$<$  2.9   &    \nodata &  0.56 & 207 &   1.12        \\
16  &  56.1814  &  32.1285  & M1       & 150 &  1.30  &$<$  2.5  &    \nodata & \nodata & \nodata & $>$   1.67; 2.25          \\
17  &  56.2339  &  32.1542  & K0    &  360 &  1.30  &$<$ 1.1   &    \nodata & \nodata & \nodata & $>$  2.15; 3.61           \\
18  &  56.2559  &  32.0838  & A3 &  -1  & \nodata  &   \nodata &    \nodata &  0.65&  271&    2.89  \\
19  &  62.6755  &  38.1000  & A3 &  -1  & 1.30   &$<$   3.9  &    \nodata & \nodata & \nodata & $>$  2.65; 4.59           \\
20  &  64.7546  &  28.3284  & M6 & 210 & 1.30  &$<$    3.3   & \nodata & \nodata & \nodata  & $>$  1.94; 3.94           \\
21  &  64.8594  &  28.4373  & K7     & 180  & 1.30  & $<$  5.4   &    \nodata & \nodata & \nodata & $>$ 2.74; 4.03           \\
22  &  65.2889  &  27.8436  & M4     & 280  &  1.30  &$<$  3.0   &    \nodata &  0.77 & 230 &     1.28      \\
23  &  66.0967  &  26.8357  & M2    & 200  &  \nodata &  \nodata   &    \nodata & \nodata & \nodata & $>$ 1.93; 3.19            \\
24  &  67.1776  &  27.2344  &  M4    & 300  &  1.30  &$<$   3.0   &    \nodata & 0.64 &  10 &    0.94     \\
25  &  67.3368  &  27.7021  &  AGB     &  -1  & \nodata   &   \nodata & \nodata & \nodata & \nodata & $>$     2.27: 3.72        \\
26  &  67.3738  &  26.2814  &  M5    & 190 & 0.85  & 4.5    &    1.1 &  0.05 &  45 &    $\sim $0     \\
27  &  67.4805  &  22.9828  &  AGB     &  -1  & 1.30   &$<$  6.0  &    \nodata & \nodata & \nodata & $>$     3.23; 3.99          \\
28  &  67.5048  &  35.2902  &  K7       &   370  &1.30   & $<$  3.3    &    \nodata & 1.3 & 24 &   1.01       \\
29  &  67.5177  &  35.3733  &  M0   &   370  & 1.30  & 9.7    &  1.5  & \nodata & \nodata & $>$  2.12; 3.67            \\
30  &  67.5685  &  35.4227  &  A4 &  -1  & 1.30  &  $<$  3.3  &    \nodata & \nodata & \nodata & $>$      2.41; 4.29       \\
31  &  67.6348  &  35.6037  &  K7   &   350  & 0.85  &  10.0   &    2.0 & 0.83 &  61 &      2.30     \\
32  &  67.6669  &  35.7029  &  K7   &   310  &  1.30  &$<$   2.5  &    \nodata &  1.2 & 307 &      $\sim$ 1   \\
33  &  67.7141  &  24.7062  &   M3    &   330  &  0.85  &$<$  8.0   &    \nodata & 0.07 & 315 &   $\sim$ 0       \\
34  &  67.8381  &  26.9784  &   AGB       &  -1  &  1.30  &$<$  5.0   &    \nodata & \nodata & \nodata & $>$  1.69; 2.66            \\
35  &  67.9377  &  28.9856  &    F5          &  -1     &  \nodata  & \nodata    &    \nodata & \nodata & \nodata & $>$   3.41; 4.94         \\
36  &  68.2676  &  29.3639  &    B9&  -1    &  1.30  &$<$ 2.8   &    \nodata & \nodata & \nodata & $>$   2.07; 3.88          \\
37  &  68.6479  &  26.7351  &    AGB      &  -1  &   0.85 &   $<$ 30.0 & \nodata & \nodata &  \nodata &  $>$   2.06; 4.33         \\
38  &  69.2047  &  24.2163  &     F2     &  -1     &   0.85  &  $<$ 27.0   &    \nodata & \nodata & \nodata & $>$   2.73; 3.96         \\
39  &  69.3536  &  27.1554  &    AGB      &  -1  &  1.30  &$<$  3.3    &    \nodata & \nodata & \nodata & $>$  3.11; 5.41           \\
40  &  69.7428  &  26.5190  &    AGB     &  -1  &     0.85&   $<$ 30.0  & \nodata & \nodata & \nodata & $>$  1.80; 3.66           \\
41  &  70.1658  &  25.3184  &    M5 &  130  & 1.30   & $<$ 2.9   &    \nodata & \nodata & \nodata & $>$  1.74; 2.34           
 \enddata
\tablenotetext{a}{``-1" implies that H$\alpha$  is seen in absorption.}
\tablenotetext{b}{$\ge$3-$\sigma$ detections or  3-$\sigma$ upper limits}
\end{deluxetable}

\begin{deluxetable}{lrrrrrrrc}
\label{derived}
\tabletypesize{\scriptsize}
\tablewidth{0pt}
\tablecaption{Derived Properties}
\tablehead{\colhead{\#}&\colhead{Log(Acc. rate)\tablenotemark{a}}&\colhead{Mass Disk\tablenotemark{b}}&\colhead{r$_{proj.}$}&\colhead{$\lambda_{tun-off}$\tablenotemark{c}}&\colhead{$\alpha_{excess}$}&\colhead{$Log$(L$_{D}$/L$_{*}$)\tablenotemark{d}}&
\colhead{$A_J$}&\colhead{Object  Type}\\
\colhead{}&\colhead{(M$_{\odot}$/yr)}&\colhead{(M$_{JUP}$)}&\colhead{(AU)}&\colhead{$\mu$m}&\colhead{}&\colhead{}&\colhead{(mag)}&\colhead{}
}
\startdata
1 & $<$ -11.0     &  $<$ 2.6  &62; 425& 5.8 & -1.19   &   -2.89  &  0.6 &  photoevaporating disk        \\
 2 &   -10.2          &          5.6 & \nodata & 5.8 &  0.69   &   -2.33  &  1.3 &  giant planet-forming disk    \\
 3 &   -9.6            &   $<$ 2.7 & \nodata & 3.6 & -0.95   &   -1.72  &  0.1 &  grain growth-dominated disk          \\
 4 &  $<$ -11.0   &  $<$ 2.6 &  240       & 8.0 & -1.69   & $<$   -3.57  &  0.5 &  debris disk          \\
 5 &   -8.5            &      5.3    &   212      & 5.8 & -0.80   &   -2.56  &  0.5 &  grain growth-dominated disk          \\
 6 &  $<$ -11.0   &  $<$ 6.2 & \nodata & 5.8 & -1.68   &   -4.27  &  0.8 &  debris disk            \\
 7 & $<$ -11.0    &  $<$ 2.7 & \nodata & 5.8 &  0.11   &   -3.09  &  0.1 &  photoevaporating disk        \\ 
 8 &  $<$ -11.0   & $<$6.2  & \nodata & 5.8 & -1.74   &   -4.09  &  0.8 &   debris disk          \\
 9 &   -10.2          &       13.0        & \nodata & 5.8 &  0.49   &   -2.19  &  0.7 &  giant planet-forming disk      \\
10 & -10.9           & \nodata & \nodata & 5.8 & -2.12   &   -4.39  &  0.4 &  debris disk           \\
11 &  $<$ -11.0  & \nodata & \nodata & 8.0 & -1.87   & $<$   -3.40  &  0.1 &   debris disk           \\
12 &  $<$ -11.0  &  $<$ 2.70 & \nodata & 8.0 & -0.20   & $<$   -2.41  &  0.5 &  photoevaporating disk          \\
13 &  $<$ -11.0  &  $<$ 1.3 & \nodata & 5.8 &  0.33   &   -2.92  &  1.5 &  photoevaporating disk           \\
14 &  $<$ -11.0  &  $<$ 2.5 &  327      & 3.6 & -0.76   &   -2.37  &  0.5 &  photoevaporating disk          \\
15 &  $<$ -11.0  &  $<$ 2.6 &   140     & 4.5 & -1.75   &   -2.80  &  1.0 &  photoevaporating disk        \\
16 &  $<$ -11.0 &   $<$ 2.2 & \nodata & 8.0 & -0.95   &  $<$ -3.03  &  2.0 &  debris disk         \\
17 & -9.4            &    $<$ 1.0 & \nodata & 4.5 & -0.35   &   -2.26  &  1.2 &  giant planet-forming disk      \\
18 & $<$ -11.0  & \nodata &   162      & 8.0 & -2.03   &  $<$  -4.05  &  0.5 &  debris disk           \\
19 &  $<$ -11.0  &  $<$ 3.0 & \nodata & 8.0 & -1.63   &  $<$  -4.88  &  0.4 &  debris disk           \\
20 &  -10.85  &  $<$ 0.6 & \nodata & 8.0 &  0.83   &   -2.14  &  0.6 &  giant planet-forming disk           \\  
21 &  $<$ -11.0 &   $<$ 0.9 & \nodata & 8.0 & -1.55   & $<$  -3.90  &  0.5 &  debris disk           \\
22 &   -10.3         &  $<$ 0.50 & 108       & 5.8 & -1.42   &   -2.57  &  0.0 &  grain growth-dominated disk       \\
23 &   $<$ -11.0 & \nodata & \nodata & 8.0 & -1.91   &   $<$ -3.78  &  0.4 &  debris disk          \\
24 &   -10.0         &  $<$ 0.50 & 90          & 4.5 & -1.18   &   -2.33  &  1.2 &  grain growth-dominated disk       \\
25 & \nodata       & \nodata & \nodata & 5.8 & -0.95   &   \nodata &  \nodata &  AGB star           \\
26 &  $<$ -11.0  &  0.4        &  7             & 8.0 & -1.88   & $<$  -3.77  &  0.0 & photoevaporating/circumbinary disk       \\
27 &  $<$ -11.0  &   $<$ 1.0 & \nodata & 8.0 & -1.97   & \nodata  &  \nodata &  AGB star           \\
28 &  -9.3            & $<$ 2.6&  390      & 8.0 & -0.90   &  $<$ -2.53  &  1.1 &  grain growth-dominated disk        \\
29 &  -9.3            &  7.5     &    \nodata               & 5.8 &  0.38   &   -2.62  &  0.7 &  giant planet-forming disk     \\
30 &  $<$ -11.0 &  $<$ 2.6 & \nodata & 8.0 & -1.39   &   $<$ -4.28  &  0.7 &  MS debris disk         \\
31 &  -9.5            &  3.7    &      249      & 5.8 &  0.90   &   -2.15  &  1.3 &   giant planet-forming disk           \\
32 &  -9.9           &  $<$ 1.9 &   360      & 5.8 & -1.05   &   -2.73  &  0.6 &  grain growth-dominated disk           \\
33 & -9.7            &  $<$ 2.9 &    10         & 5.8 & -0.99   &   -2.37  &  0.3 &  grain growth-dominated disk        \\
34 & \nodata      &  $<$ 0.8 & \nodata & 3.6 & -0.99   &  \nodata  &  \nodata &  AGB Star         \\
35 &  $<$ -11.0 & \nodata & \nodata & 8.0 & -1.32   & $<$  -3.65  &  0.7 &  MS debris disk          \\
36 &  $<$ -11.0  & $<$ 0.5 & \nodata & 5.8 &  0.09   &   -3.44  &  0.5 &  MS debris disk          \\ 
37 & \nodata      & \nodata & \nodata & 8.0 & -1.94   &  \nodata &  \nodata &  AGB Star         \\
38 &  $<$ -11.0 & \nodata & \nodata & 8.0 & -1.71   &   $<$ -3.15  &  0.0 &  MS debris disk         \\ 
39 & \nodata      & $<$ 0.6 & \nodata & 8.0 & -1.86   &   \nodata  &  \nodata &  AGB Star          \\
40 & \nodata       & \nodata & \nodata & 8.0 & -1.93   &   \nodata  &  \nodata &  AGB star          \\
41 &  $<$ -11.0  & $<$ 0.5 & \nodata & 8.0 & -1.35   &   $<$ -3.90  &  0.9 &  debris disk         
\enddata
\tablenotetext{a}{~The uncertainties in the accretion rate are dominated by the calibration of Equation. The values  should be considered order of magnitude
estimates. See Section 4.2.}
\tablenotetext{b}{~Disk masses were derived using Equations 2 and 3, and should be within a factor of $\sim$2 of modeled derived masses; however, 
larger systematic errors cannot be ruled out due to uncertainties in dust opacities and  gas to dust mass ratios. See Section 4.3.}
\tablenotetext{c}{~The uncertainty in $\lambda_{tun-off}$ is 1 SED point.  See Section 4.5}
\tablenotetext{d}{~$log$(L$_{D}$/L$_{*}$) is highly dependent on the SED sampling and should be consider an order of magnitude estimate. See Section 4.5.}
\end{deluxetable}

\begin{deluxetable}{lcccc}
\label{incidence}
\tabletypesize{\scriptsize}
\tablewidth{0pt}
\tablecaption{Disk types  as a function of spectral type for combined sample\tablenotemark{a}}
\tablehead{\colhead{Spectral types}&\colhead{Grain-growth dominated}&\colhead{Giant planet-forming}&\colhead{Photoevaporating}&\colhead{Debris}
}
\startdata
All  stars                         & 39.2$\%$    (29/74)  & 17.6$\%$  (13/74)  &    20.3$\%$ (15/74)  &     23.0$\%$ (17/74) \\
M-type stars                  & 39.6$\%$   (19/48)  & 16.6$\%$    (8/48)   &    25.0$\%$ (12/48)   &    18.8$\%$ (9/48)      \\
K-type stars                   & 58.8$\%$     (10/17)  &  29.4$\%$(5/17)   &    0.0$\%$  (0/17)    &    11.8$\%$  (2/17)  \\
GAFB-type   stars         &0.0$\%$           (0/9)   &  0.0$\%$   (0/9)   &     33.3$\%$  (3/9)     &      66.6$\%$  (6/9)    \\
\enddata
\tablenotetext{a}{It includes all transition disks from Papers I, II, and III. 
}
\end{deluxetable}

\begin{deluxetable}{rcccc}
\label{incidence}
\tabletypesize{\scriptsize}
\tablewidth{0pt}
\tablecaption{Disk types as a function of  age for combined sample\tablenotemark{a}}
\tablehead{\colhead{Age}&\colhead{Grain-growth dominated}&\colhead{Giant planet-forming}&\colhead{Photoevaporating}&\colhead{Debris}
}
\startdata
$<$ 1 Myr                    & 71.4$\%$    (10/14)  &  0.0$\%$    (0/14)  &    14.3$\%$ (2/14)  &    14.3$\%$ (2/14)  \\  
1-3 Myr                        & 33.3$\%$    (10/30)  & 26.7$\%$   (8/30)  &    20.0$\%$ (6/30)  &     20.0$\%$ (6/30) \\
$>$ 3 Myr                     & 30.0$\%$     (9/30)   & 16.7$\%$   (5/30)  &    23.3$\%$ (7/30)   &    30.0$\%$  (9/30) \\ 
\enddata
\tablenotetext{a}{It includes all transition disks from Papers I, II, and III. 
}
\end{deluxetable}

\begin{figure}[th]
\includegraphics[width=5.5in]{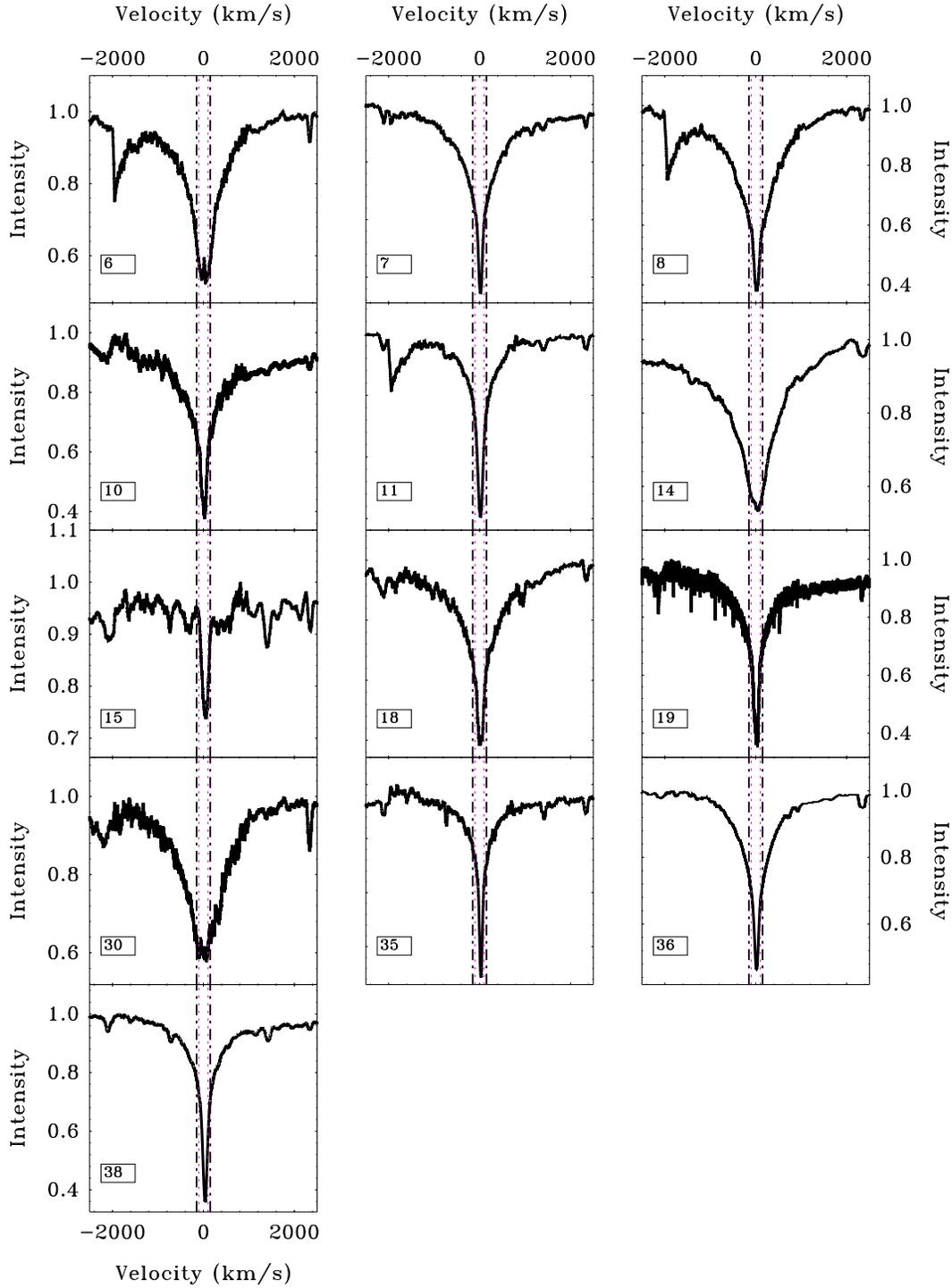}
\caption{
The H$\alpha$ profiles of the 13 non-accreting 
objects in our sample where H$\alpha$ was detected in absorption.
They are all GFAB-type stars.
}
\label{f:prof_abs}
\end{figure}

\end{document}